\documentclass[aps,pra,twocolumn,superscriptaddress,nofootinbib]{revtex4-2}

\usepackage{amsmath,amssymb,amsfonts,bm}
\usepackage{verbatim}
\usepackage{booktabs}
\usepackage{graphicx, xcolor}
\usepackage{subcaption}
\usepackage{physics, braket}
\usepackage{quantikz}
\usepackage{hyperref}

\renewcommand{\Tr}{\mathrm{Tr}}

\begin{document}

\title{Qubit-Boson Hybrid Beam-Splitter Gate with Kerr Nonlinearity in Circuit QED for Many-Body Dynamics}
\author{A.~Mammola}
\affiliation{C12 Quantum Electronics, Paris, France}
\affiliation{Aix Marseille Univ, CNRS, LIS, Marseille, France}
\author{G.~Di Molfetta}
\affiliation{Aix Marseille Univ, CNRS, LIS, Marseille, France}
\affiliation{Institut Universitaire de France, Paris, France}
\author{Q.~Schaeverbeke }
\affiliation{C12 Quantum Electronics, Paris, France}
\date{\today}
\begin{abstract}
We introduce a hybrid qubit-boson beam-splitter gate in which a microwave cavity mode couples to an exchange-dressed two-level subsystem of an interacting two-qubit system in the presence of Kerr nonlinearity. Starting from a general circuit quantum electrodynamics (cQED) model, we derive the corresponding hybrid qubit-cavity interaction, develop its open-system description including photon- and qubit-sector-bath-induced dissipation and obtain in the weak-dissipation regime an analytical expression for the average gate fidelity.
%
%
We further identify carbon-nanotube circuit QED as a concrete platform for implementing and controlling the gate, provide a representative operating regime, and perform noiseless and noisy numerical simulations to study the gate dynamics and benchmark the analytical results.
Beyond this implementation route, the proposed hybrid primitive provides a natural building block for many-body dynamics, including quantum-cellular-automaton (QCA) and lattice-gauge-inspired architectures and, through its collision-model reformulation, also suggests connections to noisy QCA, non-Markovian extensions and reservoir-style quantum information processing.
\end{abstract}
%

\maketitle

\section{Introduction}

A central challenge in quantum information processing is to identify physically realistic settings in which useful gate primitives can be implemented natively, rather than synthesized from long sequences of elementary operations. This is particularly important in the current and near-term NISQ era \cite{Preskill}, where exploiting the natural interaction structure of a platform can substantially reduce overhead and enable digital-semi-analog approaches to quantum dynamics. More generally, the same hardware-aware viewpoint remains relevant in the longer term, for instance in quantum simulation and quantum error correction, where the form of the available couplings strongly affects both scalability and resource efficiency.

In this respect, exchange-type interactions of Heisenberg or XY form \cite{XY} are especially attractive, since they naturally generate excitation-preserving gates such as iSWAP and related partial-SWAP operations. Such gates are broadly useful: they provide natural primitives for the simulation of spin models and transport dynamics \cite{GeorgescuQuantumSimulation}, underlie Hamming-weight-preserving circuits relevant to quantum machine learning and constrained state preparation \cite{MonbroussouQMLHWP,ShaydulinXYMixer} and play an important role in reconfigurable architectures for quantum error correction \cite{EickbuschDynamicSurfaceCodes}. From a hardware perspective, such gates have been extensively studied as native entangling interactions emerging from either nearest-neighbor Heisenberg-like couplings or long-range cavity-mediated interactions, for instance in superconducting and semiconducting-spin platforms \cite{AbramsXYGate,DijkemaCavityISWAP}.

The present work explores a different use of the bosonic mode and exchange interaction. Rather than operating the cavity only as a virtual mediator in a dispersive regime, we consider a genuinely hybrid setting in which a cavity degree of freedom participates actively in the dynamics. More precisely, we combine qubit--qubit exchange with longitudinal qubit--cavity coupling, so that the odd-parity sector of an interacting two-qubit manifold is coupled directly to cavity number-changing transitions. In this way, the exchange interaction dresses the odd-parity sector into an effective two-level subsystem, whose coupling to a microwave cavity mode yields what we call a hybrid qubit-boson beam-splitter gate. This interaction can then be composed with Kerr evolution to generate number-dependent phases.

This leads to a parity-preserving hybrid qubit-boson dynamics that extends the usual exchange-gate paradigm beyond simple virtual-photon mediation and into a larger Hilbert space. In this way, the controllability of few-level qubit systems is enriched by the larger bosonic Hilbert space of the cavity, where occupation-dependent processing and nonlinear phases can play an active role. Accordingly, in this work we derive the resulting beam-splitter gate from a general circuit-QED effective model, analyze its coherent dynamics and then develop its open-system description.

A complementary aspect of this work is to identify a concrete physical setting in which the same ingredients can be realized and controlled. In this respect, carbon-nanotube circuit QED is particularly promising \cite{neukelmance2025microsecond}. Carbon nanotubes naturally support double-quantum-dot architectures with electrically tunable exchange interactions, transversal and longitudinal spin-photon coupling to microwave resonators allowing for high connectivity \cite{MammolaPRA2025} and flexible microwave control, all within a scalable solid-state setting \cite{cottet2010spin,penfold2017microwave,delbecq2020,khivrich2020atomic,chen2023long}. They are also appealing from the viewpoint of hardware-oriented quantum cellular automata (QCA) \cite{Arrighi}, as noisy quantum-walk dynamics implemented through QCA on semiconducting spin-processor architectures have already been studied and benchmarked on noisy emulators \cite{wc4h-thmg}. This makes them a natural realization route for the hybrid gate studied here and for the broader many-body perspectives via QCA discussed later in the paper. 

Indeed, beyond its direct gate interpretation, the present primitive is motivated by several broader directions. First, it provides a natural building block for locally updated hybrid many-body dynamics, including quantum-cellular-automaton and lattice-gauge-inspired constructions \cite{SellapillayArrighiDiMolfettaQCAQED}, in which a persistent bosonic mode mediates matter-field exchange and carries local field information across successive updates. Second, the same setup admits a collision-model reformulation \cite{CiccarelloCollisionReview,CusumanoCollisionGuide}, from which one recovers the same effective dissipative structure and which naturally connects the gate to ancilla-based open-system simulation. Finally, the combination of a parity-preserving qubit sector, an actively participating bosonic mode, Kerr-induced nonlinearity and structured dissipation also points toward richer forms of nonclassical and memory-bearing dynamics, with possible links to noisy QCA \cite{AvalleSerafiniNoisyQCA} and reservoir-style quantum information processing \cite{SanniaDissipativeQRC}.

Our main contributions are threefold.
\begin{enumerate}
    \item First, we formulate and analyze a general effective model for a hybrid qubit-boson beam-splitter gate, and derive the corresponding effective qubit-cavity exchange interaction together with its composition with Kerr evolution.

    \item Second, we develop a Markovian open-system description \cite{BreuerPetruccione,Davies1974} including photon- and qubit-sector-bath-induced dissipation and show that the same effective dissipative dynamics can be obtained both from the standard Born--Markov--secular treatment and from a repeated-interaction collision-model construction. In the weak-dissipation regime, we further derive a compact analytical expression for the average gate fidelity, benchmarked against simulations performed with the Python-based QuTiP library \cite{Johansson2012QuTiP,Johansson2013QuTiP}.

    \item Third, we identify carbon-nanotube circuit QED as a concrete platform in which the ingredients required for the gate can be implemented and controlled, and we quantify a representative operating regime compatible with the approximations underlying the effective and open-system treatments.
\end{enumerate}

The paper is organized as follows. Section~\ref{sec:ideal} introduces the general circuit-QED effective model and derives the hybrid beam-splitter interaction together with the Kerr nonlinearity. Section~\ref{sec:lindblad_main} presents the open-system dynamics and derives the same effective master equation using both standard and collision-model methods. Section~\ref{sec:fidelity_main} discusses the average gate fidelity in the weak-dissipation regime. Section~\ref{sec:cnt_design_recipe} introduces the carbon-nanotube circuit-QED qubit architecture, presents a concrete realization route for the gate, and identifies a representative operating regime based on realistic parameters. Section~\ref{sec:simulations} presents numerical simulations of the noiseless and noisy dynamics and benchmarks the analytical results. Finally, Section~\ref{sec:conclusion} summarizes the main results and discusses broader perspectives toward hybrid many-body dynamics, open-system simulation, and nonlinear qubit-boson processing.

\begin{figure*}[t]
\centering
\includegraphics[width=\textwidth]{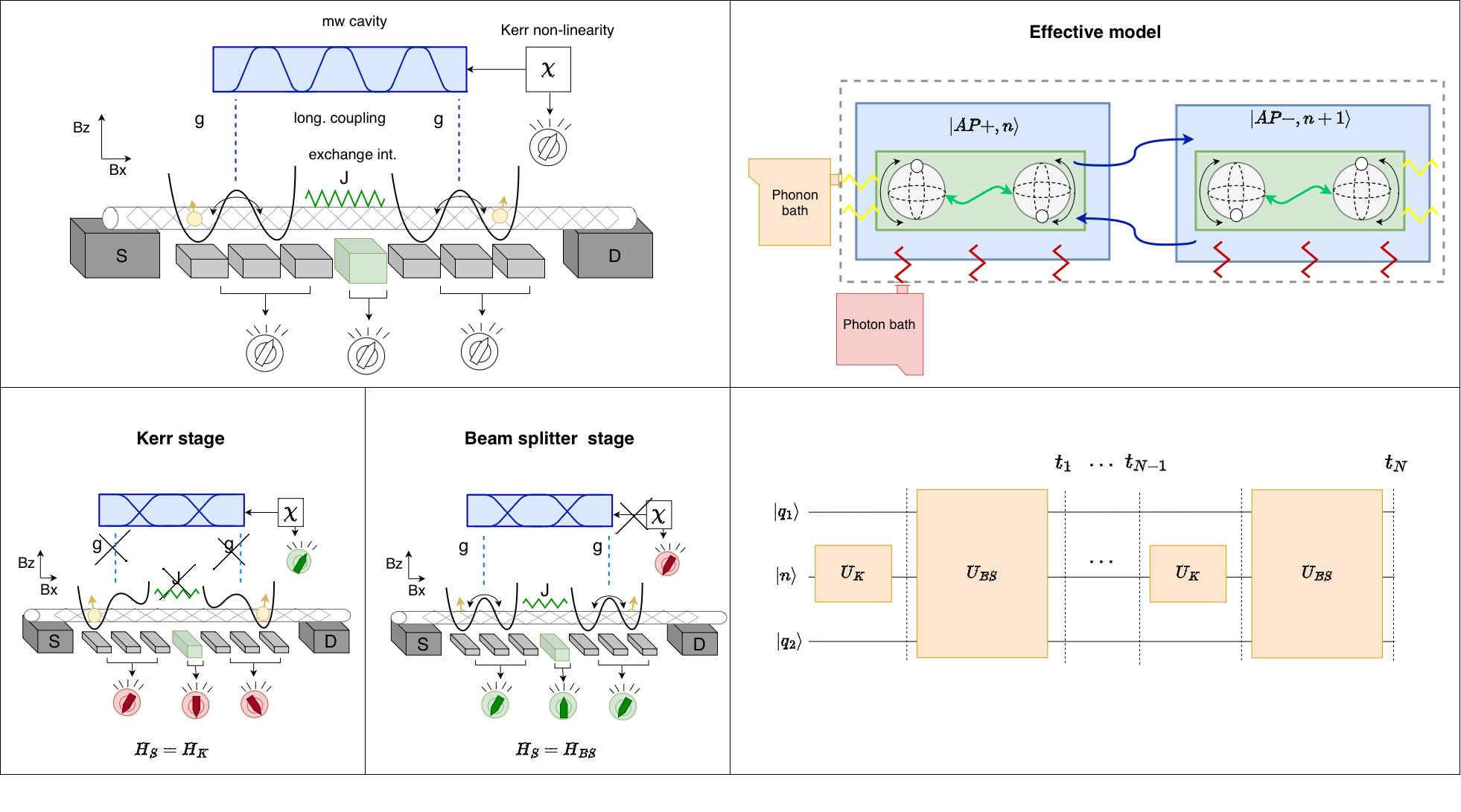}
\caption{Pseudo-3D schematic of a carbon-nanotube circuit-QED realization of the hybrid gate. Top left: experimental platform, where a suspended CNT above electrostatic gates defines two double quantum dots with tunable exchange interaction $J$, both longitudinally coupled to a microwave resonator, with a weak tunable Kerr nonlinearity $\chi$ engineered at the cavity level. Top right: effective dressed-state description in the odd-parity sector, where the states $\{\ket{AP+},\ket{AP-}\}$ couple to the cavity mode through an effective beam-splitter interaction, while the cavity also acquires a weak Kerr nonlinearity and the environment induces phonon- and photon-sector dissipation. Bottom left and bottom center: operating configurations corresponding to the Kerr and beam-splitter stages of the protocol. Bottom right: circuit representation of the stroboscopic sequence, alternating Kerr gates $U_K$ and beam-splitter gates $U_{\mathrm{BS}}$.}
\label{fig:cnt_design}
\end{figure*}

\section{Effective Hamiltonian and ideal gate}
\label{sec:ideal}

\subsection{Microscopic model}

We consider a generic hybrid circuit-QED setting in which two capacitively gate defined double quantum dots are coupled longitudinally coupled to a single microwave cavity mode. The minimal ingredients needed for the gate studied here are: (i) a qubit--qubit exchange interaction, (ii) a longitudinal qubit--cavity coupling, and (iii) a weak cavity nonlinearity. At this level, the construction is not tied to a unique microscopic platform, however a concrete implementation based on carbon nanotubes is discussed below.
With these ingredients, we consider the system Hamiltonian
\begin{equation}
H_{S}=H_{Q}+H_{c},
\label{eq:H_full_main}
\end{equation}
with
\begin{equation}
H_{Q}=
\sum_{i=1}^2 \frac{\omega_i}{2}\sigma_z^{(i)}
+ g(\sigma_z^{(1)}-\sigma_z^{(2)})(a+a^\dagger)
+ J\sigma_x^{(1)}\sigma_x^{(2)},
\label{eq:H_Q_restructured}
\end{equation}
and
\begin{equation}
H_{c}=\omega_c a^\dagger a+\chi (a^\dagger a)^2,
\label{eq:H_c_restructured}
\end{equation}
where $\omega_{1,2}$ are the qubit splittings, $g$ is the longitudinal qubit--cavity coupling strength, $J$ is an exchange-like qubit--qubit interaction, $a$ is the cavity annihilation operator, $\omega_c$ is the cavity frequency and $\chi$ is the Kerr coefficient. The exchange term mixes even (parallel) and odd (antiparallel) parity qubit configurations, while the longitudinal coupling dresses these manifolds through the cavity mode.
The Kerr term in Eq.~\eqref{eq:H_c_restructured} is introduced as an effective weak cavity nonlinearity, generating number-dependent phases that can be composed with the beam-splitter dynamics. Its motivation is simulation-driven, e.g. lattice gauge simulations via QCA, and will be discussed in section ~\ref{sec:conclusion}.
\subsection{Effective beam-splitter interaction}

Let us now derive the targeted hybrid beam-splitter interaction. See Appendix~\ref{app:supp_effham} for the detailed derivation.
We start by applying the polaron transformation
\begin{equation}
U_P=\exp[\lambda Z(a^\dagger-a)],
\qquad
Z=\sigma_z^{(1)}-\sigma_z^{(2)},
\qquad
\lambda=\frac{g}{\omega_c}.
\end{equation}
which, exactly removes the longitudinal qubit--cavity coupling and, to leading order in $\lambda$, yields the transformed qubit Hamiltonian
\begin{equation}
\begin{split}
H_{Q}^{P} &:=U_{P}^{\dagger}H_{Q}U_{P}\\
&\approx
\sum_{i=1}^2\frac{\omega_i}{2}\sigma_z^{(i)}
+\frac{2g^2}{\omega_c}\sigma_z^{(1)}\sigma_z^{(2)}
+J\sigma_x^{(1)}\sigma_x^{(2)}
+H_{J},
\end{split}
\label{eq:H_polaron_restructured}
\end{equation}
where
\begin{equation}
\label{eq:polaron_exchange}
H_{J}=
4J\lambda(a^\dagger-a)
\Big(\sigma_+^{(1)}\sigma_-^{(2)}-\sigma_-^{(1)}\sigma_+^{(2)}\Big).
\end{equation}
For the cavity Hamiltonian we assume $\chi\ll\omega_c$ and restrict to low photon numbers, so that $U_{P}^{\dagger}H_{c}U_{P}\approx H_c$.

The exchange term $\sigma_x^{(1)}\sigma_x^{(2)}$ couples $\ket{00}\leftrightarrow\ket{11}$ and $\ket{01}\leftrightarrow\ket{10}$, so that the qubit Hamiltonian separates into two independent sectors: the even-parity (parallel) manifold $\mathcal H_{\rm p}=\mathrm{span}\{\ket{00},\ket{11}\}$ and the odd-parity (anti-parallel) manifold $\mathcal H_{\rm ap}=\mathrm{span}\{\ket{01},\ket{10}\}$. We diagonalize them independently.

\paragraph{Even parity manifold.}
In the even-parity sector we define
\begin{align}
\ket{\mathrm{P}+}&=\cos\theta_{\rm p}\ket{00}+\sin\theta_{\rm p}\ket{11},\\
\ket{\mathrm{P}-}&=-\sin\theta_{\rm p}\ket{00}+\cos\theta_{\rm p}\ket{11},
\end{align}
with
\begin{equation}
\tan(2\theta_{\rm p})=\frac{2J}{\omega_1+\omega_2},
\qquad
\Omega_{\rm p}=\sqrt{J^2+\big[(\omega_1+\omega_2)/2\big]^2}.
\end{equation}
In this sector, the diagonal Hamiltonian is
\begin{equation}
\mathcal{H}_{\rm p}^{\prime}=
\frac{2g^2}{\omega_c}\,\mathbb I_{\rm p}
+\Omega_{\rm p}
\Big(
\ket{\mathrm{P}+}\!\bra{\mathrm{P}+}
-
\ket{\mathrm{P}-}\!\bra{\mathrm{P}-}
\Big).
\end{equation}

\paragraph{Odd parity manifold.}
In the odd-parity sector we define
\begin{equation}
\begin{split}
|AP+\rangle &= \cos\theta_{\mathrm{ap}} |01\rangle 
+ \sin\theta_{\mathrm{ap}} |10\rangle,
\\
|AP-\rangle &= -\sin\theta_{\mathrm{ap}} |01\rangle 
+ \cos\theta_{\mathrm{ap}} |10\rangle,
\end{split}
\end{equation}
with
\begin{equation}
\tan(2\theta_{\mathrm{ap}})=\frac{2J}{\Delta},
\quad
\Delta=\omega_1-\omega_2,
\quad
\Omega_{\mathrm{ap}}=\sqrt{\left(\frac{\Delta}{2}\right)^2+J^2},
\end{equation}
and energies
\begin{equation}
E_{\mathrm{AP}\pm}=-\frac{2g^2}{\omega_c}\pm\Omega_{\rm ap},
\qquad
\omega_{\rm ap}=E_{\mathrm{AP}+}-E_{\mathrm{AP}-}=2\Omega_{\rm ap}.
\end{equation}
Introducing Pauli operators in this dressed odd-parity subspace,
\begin{align}
\tau_z&=\ket{\mathrm{AP}+}\!\bra{\mathrm{AP}+}-\ket{\mathrm{AP}-}\!\bra{\mathrm{AP}-},\\
\tau_+&=\ket{\mathrm{AP}+}\!\bra{\mathrm{AP}-},\qquad
\tau_-=\ket{\mathrm{AP}-}\!\bra{\mathrm{AP}+},
\end{align}
the effective odd-parity qubit Hamiltonian reads
\begin{equation}
\mathcal{H}_{\rm ap}^{\prime}=
-\frac{2g^2}{\omega_c}\,\mathbb I_{\rm ap}
+\Omega_{\rm ap}\tau_z,
\end{equation}
while the polaron-induced cavity-assisted exchange becomes
\begin{equation}
\mathcal{H}_{\rm ap,J}^{\prime}=
4iJ\lambda(a^\dagger-a)(\tau_- - \tau_+).
\label{eq:HJ_tauy_restructured}
\end{equation}

Moving to the interaction picture with respect to
\begin{equation}
H_0=\omega_c a^\dagger a+\mathcal H_{\rm ap}^{\prime},
\end{equation}
and working near the resonance condition $\omega_c\simeq\omega_{\rm ap}$, the interaction Hamiltonian becomes $H_{BS}+H_{K}$, with

\begin{equation}
H_{BS}(t)
\simeq
g_{\rm eff}
\Big(
a^\dagger\tau_- e^{i\delta t}
+
a\tau_+ e^{-i\delta t}
\Big),
\quad
g_{\rm eff}=4J\frac{g}{\omega_{c}},
\label{eq:H_rsb_restructured}
\end{equation}
where $\delta=\omega_c-\omega_{\rm ap}$ is the detuning and $H_{K}:=\chi(a^{\dagger}a)^{2}$ the Kerr Hamiltonian. The resulting interaction is of Jaynes--Cummings type, with the effective two-level system given by the dressed odd-parity manifold. In this sense, it defines a hybrid qubit-boson beam-splitter interaction between the cavity mode and the dressed states $\{\ket{AP-},\ket{AP+}\}$. 
\subsection{Gate construction}
\label{sec:gate_construction}

The resulting gate is a composition of a hybrid qubit-boson beam-splitter evolution and a Kerr phase evolution,
\begin{equation}
U=e^{-it(H_{BS}+H_{K})},
\end{equation}
which, because $[H_{BS}(t),H_K]\neq 0$, does not in general factorize exactly into two separate contributions. If the Kerr phase accumulated during the beam-splitter step remains small over the occupied manifold, i.e.\ $|\chi|\,t_g\,n_{\max}\ll 1$, the dynamics can nevertheless be treated perturbatively through a first-order Zassenhaus expansion \cite{Casas2012Zassenhaus}. More generally, depending on the available control resources, the beam-splitter and Kerr terms may also be implemented sequentially, in separate control windows. In the following, we restrict to this case, for which
\begin{equation}
U=U_{BS}\,U_K,
\qquad
U_K=\exp\!\big[-it_K \chi (a^\dagger a)^2\big],
\label{eq:U_seq}
\end{equation}
where, in the dressed diagonal qubit basis
$\{\ket{P+},\ket{AP-},\ket{AP+},\ket{P-}\}$ and at exact resonance
$\delta=0$, the beam-splitter evolution generated by the effective exchange Hamiltonian takes the form
\begin{equation}
\begin{split}
U_{BS}(t_{BS})
&=e^{-iH_{BS}t_{BS}}
\\
&=
\begin{pmatrix}
\mathbb I & 0 & 0 & 0\\[2mm]
0 &
\cos\!\big(\Theta_{\hat N}\big) &
-i\,a^\dagger\,
\dfrac{\sin\!\big(\Theta_{\hat N+1}\big)}{\sqrt{\hat N+1}}
& 0\\[4mm]
0 &
-i\,a\,
\dfrac{\sin\!\big(\Theta_{\hat N}\big)}{\sqrt{\hat N}} &
\cos\!\big(\Theta_{\hat N+1}\big)
& 0\\[4mm]
0 & 0 & 0 & \mathbb I
\end{pmatrix},
\end{split}
\label{eq:UBS_operator_main}
\end{equation}
where $\hat N=a^\dagger a$ and
\begin{equation}
\Theta_{\hat N}=g_{\mathrm{eff}} t_{BS}\sqrt{\hat N},
\qquad
\Theta_{\hat N+1}=g_{\mathrm{eff}} t_{BS}\sqrt{\hat N+1},
\end{equation}
$\mathbb I$ denotes the identity on the cavity Hilbert space and
$t_{BS}$ is the beam-splitter interaction time associated with the effective coupling $g_{\rm eff}=4Jg/\omega_c$.

This expression should be understood blockwise: the interaction acts nontrivially only within the dressed odd-parity manifold $\{\ket{AP\pm}\}$, while the even-parity states remain spectators. More precisely, the odd-parity dynamics decomposes into invariant subspaces spanned by $\{\ket{AP+,n},\ket{AP-,n+1}\}$, each undergoing a two-level Rabi rotation with angle $g_{\mathrm{eff}}t_{BS}\sqrt{n+1}$. A detailed derivation of this block structure is given in the Supplemental Material.

\section{Open-system dynamics}
\label{sec:lindblad_main}
We now derive the effective Markovian master equation governing the dressed qubit-cavity dynamics in the presence of cavity and matter-sector dissipation. Concretely, we consider a photon bath coupled to the cavity and a second bath coupled to the qubit sector, generating generic relaxation and dephasing channels. In the carbon-nanotube circuit-QED realization discussed in Sec.~\ref{sec:cnt_design_recipe}, the latter is naturally identified with a phonon environment.

As follows, we present two complementary derivations. The first is the standard Born--Markov--Davies derivation \cite{BreuerPetruccione,Davies1974} applied to the effective dressed system-bath couplings. The second reformulates the same open-system dynamics as a repeated-interaction, or collision-model and derives the same GKLS generator in the weak-collision continuum limit \cite{CiccarelloCollisionReview,CusumanoCollisionGuide}. See Appendix~\ref{app:supp_lindblad} for a detailed derivation.

\subsection{System--bath Hamiltonian}

We model the environment as the combination of a photon bath and a phonon bath coupled to the qubits+cavity system. The total Hamiltonian is
\begin{equation}
H_{\mathrm{tot}}=H_S+H_\gamma+H_{\gamma S}+H_\nu+H_{\nu S},
\end{equation}
where $H_S$ is the system Hamiltonian of Eq.~\eqref{eq:H_full_main}. The photon bath and its coupling are
\begin{align}
H_\gamma&=\sum_k \omega_k b_k^\dagger b_k,\\
H_{\gamma S}&=S_{\gamma}\otimes B_\gamma,\qquad 
B_\gamma=\sum_k (f_k b_k^\dagger+f_k^\ast b_k),\\
S_{\gamma}&=a+a^\dagger,
\end{align}
while the phonon bath is described by
\begin{align}
H_\nu&=\sum_q \omega_q c_q^\dagger c_q,\\
H_{\nu S}&=S_\nu\otimes B_\nu,\qquad 
B_\nu=\sum_q (h_q c_q^\dagger+h_q^\ast c_q),\\
S_\nu&=\sum_{i=1}^2\left(\alpha_i\sigma_x^{(i)}+\beta_i\sigma_z^{(i)}\right),
\label{eq:Snu_def_main_new}
\end{align}
where $\alpha_i$ parametrizes transverse (flip) coupling and $\beta_i$ longitudinal (dephasing-type) coupling.
For both derivations we work in the same effective dressed description as in Sec.~\ref{sec:ideal}: we apply the polaron transformation to remove the longitudinal qubit-cavity coupling to leading order and diagonalize the even- and odd-parity qubit sectors. 

For the cavity Hamiltonian, we assume \(\chi\ll\omega_c\) and restrict to low photon numbers, so that \(U_{P}^{\dagger}H_{c}U_{P}\approx H_c\).
Within this effective description, two Kerr regimes can be distinguished. In the unresolved-Kerr regime, the Kerr-induced spread of cavity transition frequencies across the occupied manifold remains small compared with the relevant bath linewidth (\(2|\chi|n_{\max}\ll\kappa\), for a cavity linewidth \(\kappa\)), so that individual Kerr-shifted transitions cannot be resolved and the cavity dissipator is well approximated by the usual collapse operators \(a\) and \(a^\dagger\). If, instead, the Kerr splitting becomes comparable to or larger than the bath linewidth, the cavity transitions become number dependent, with frequencies and jump operators
\begin{align}
\omega_n&=\omega_c+\chi(2n-1),\qquad n\ge1,\\
a_n&=\sqrt{n}\,|n-1\rangle\langle n|.
\label{eq:kerr_shift_main}
\end{align}
In this work, we focus on the unresolved-Kerr regime. In particular, this means that in a sequential implementation the cavity loss retains the same \(a\) and \(a^\dagger\) structure during both the hybrid beam-splitter and Kerr stages. For simplicity, during the Kerr stage we retain only the cavity dissipation, assuming that the effective qubit-related decoherence channels are sufficiently suppressed in the parked configuration, so that their contribution over the Kerr timescale remains subleading with respect to cavity loss. For instance, in the CNT cQED realization discussed in Sec.~\ref{sec:cnt_design_recipe}, this assumption is motivated by the fact that the qubits can be detuned toward an idle regime with reduced spin-charge hybridization, so that the phonon-sensitive matrix elements are correspondingly suppressed and their contribution over the Kerr window is expected to remain subleading with respect to cavity loss.

The polaron-induced exchange term $H_J$ of Eq.~\eqref{eq:polaron_exchange} is retained in the coherent dynamics but neglected in the dissipative eigenoperator decomposition. This is valid provided the splittings induced by $H_J$ remain unresolved by the baths, i.e.\ when $g_{\rm eff}\sqrt{n_{\max}+1}$ is small compared with the relevant bath linewidths. In that regime, the baths probe the unhybridized dressed spectrum while $H_J$ continues to act coherently. 

\subsection{Born--Markov--secular derivation}

We first derive the master equation using the standard Born--Markov--Davies decomposition. In this framework, each system operator coupled to a bath is decomposed into eigenoperators of the reference Hamiltonian,
\begin{equation}
S(\omega)=\sum_{\epsilon-\epsilon'=\omega}
\Pi(\epsilon)\,S\,\Pi(\epsilon'),
\qquad
[H_0,S(\omega)]=-\omega S(\omega),
\end{equation}
where $\Pi(\epsilon)$ projects onto the eigenspace of $H_0$ with energy $\epsilon$. The reduced density matrix then obeys the GKLS master equation
\begin{equation}
\begin{split}
\dot{\rho}
&=
-i[H(t),\rho]
+
\sum_{\omega}
\mathcal{D}[L(\omega)]\rho, \\
\label{eq:lindblad_main_revised_new}
\end{split}
\end{equation}
with dissipators $\mathcal{D}[L]\rho
=
L\rho L^\dagger
-
\frac12\{L^\dagger L,\rho\}$
and jump operators $L(\omega)=\sqrt{\gamma(\omega)}\,S(\omega)$,
where $\gamma(\omega)$ is the bath spectral rate evaluated at frequency $\omega$.

\subsubsection{Photon-induced channels}

After the polaron transformation, the photon coupling becomes
\begin{equation}
\label{eq:polaron_cavity}
S_{\gamma}^{P}:=U_{P}^{\dagger}S_{\gamma}U_{P}=a+a^\dagger-2 \lambda Z, 
\end{equation}
where $Z=\sigma_z^{(1)}-\sigma_z^{(2)}$.
In the unresolved-Kerr regime, the cavity contribution is described by the standard collapse operators
\begin{equation}
L_a=\sqrt{\kappa_\downarrow}\,a,
\qquad
L_{a^\dagger}=\sqrt{\kappa_\uparrow}\,a^\dagger,
\label{eq:cavity_jumps_new}
\end{equation}
where $\kappa_\downarrow=\gamma_\gamma(\omega_c)$ and $\kappa_\uparrow=\gamma_\gamma(-\omega_c)$ are the photon emission and absorption rates. In the resolved-Kerr regime, these channels are simply refined into the number-resolved operators given by Eqs.~\eqref{eq:kerr_shift_main}.

The polaron transformation also generates the additional qubit operator component proportional to $Z$, see Eq.(~\ref{eq:polaron_cavity}). Projecting it onto the dressed qubit eigenbasis yields effective photon-induced transitions in the odd-parity (anti-parallel) manifold $\{|AP+\rangle,|AP-\rangle\}$ with splitting $\omega_{\mathrm{ap}}=2\Omega_{\mathrm{ap}}$. The corresponding collapse operators are
\begin{equation}
\begin{split}
L_{\gamma,+}
&=
\sqrt{\gamma_\gamma(\omega_{\mathrm{ap}})}\,
(4\lambda\sin 2\theta_{\mathrm{ap}})\,
|AP-\rangle\langle AP+|,
\\
L_{\gamma,-}
&=
\sqrt{\gamma_\gamma(-\omega_{\mathrm{ap}})}\,
(4\lambda\sin 2\theta_{\mathrm{ap}})\,
|AP+\rangle\langle AP-|,
\\
L_{\gamma,0}
&=
\sqrt{\gamma_\gamma(0)}\,
(4\lambda\cos 2\theta_{\mathrm{ap}})\\
&\big(|AP+\rangle\langle AP+|
-|AP-\rangle\langle AP-|\big).
\label{eq:photon_qubit_jumps_new}
\end{split}
\end{equation}
The first two operators describe photon-induced relaxation and excitation between the dressed antiparallel states, while $L_{\gamma,0}$ generates pure dephasing in that manifold. The even-parity manifold $\{|P\pm\rangle\}$ is unaffected because $Z$ annihilates $|00\rangle$ and $|11\rangle$.

\subsubsection{Qubit-sector bath-induced channels}

In the polaron frame, the qubit-sector bath coupling operator to leading order in $\lambda$ reads 
\begin{equation}
S_\nu^{P}
:=U_{P}^{\dagger}S_{\nu}U_{P}=
S_\nu+\lambda S_\nu^{(1)}+\mathcal O(\lambda^2),
\label{eq:Snu_split_new}
\end{equation}
with
\begin{align}
S_\nu
&=
\sum_{i=1}^2\big(\alpha_i\sigma_x^{(i)}+\beta_i\sigma_z^{(i)}\big),
\label{eq:Snu0_def_new}
\\
S_\nu^{(1)}
&=
2i(a^\dagger-a)
\big(\alpha_1\sigma_y^{(1)}-\alpha_2\sigma_y^{(2)}\big).
\label{eq:Snu1_def_new}
\end{align}

Decomposing $S_\nu$ in the dressed eigenbasis produces three classes of channels.

\paragraph{Even-parity sector.}
\begin{equation}
\begin{split}
L_{\nu,{\rm p}}(\pm\omega_{\rm p})
&=
-\sqrt{\gamma_\nu(\pm\omega_{\rm p})}(\beta_1+\beta_2)\sin(2\theta_{\rm p})
\ket{P\mp}\!\bra{P\pm},\\
L_{\nu,{\rm p}}(0)
&=
\sqrt{\gamma_\nu(0)}(\beta_1+\beta_2)\cos(2\theta_{\rm p})\\
&\Big(\ket{P+}\!\bra{P+}
-\ket{P-}\!\bra{P-}\Big).
\end{split}
\end{equation}

\paragraph{Odd-parity sector.}
\begin{equation}
\begin{split}
L_{\nu,{\rm ap}}(\pm\omega_{\rm ap})
&=
-\sqrt{\gamma_\nu(\pm\omega_{\rm ap})} (\beta_1-\beta_2)\sin(2\theta_{\rm ap})\\
&\ket{\mathrm{AP}\mp}\!\bra{\mathrm{AP}\pm},\\
L_{\nu,{\rm ap}}(0)
&=
\sqrt{\gamma_\nu(0)}\,(\beta_1-\beta_2)\cos(2\theta_{\rm ap})\\
&\Big(\ket{\mathrm{AP}+}\!\bra{\mathrm{AP}+}
-\ket{\mathrm{AP}-}\!\bra{\mathrm{AP}-}\Big).
\end{split}
\end{equation}

\paragraph{Cross-manifold transitions.}
Single-qubit flip terms proportional to $\sigma_x^{(i)}$ induce transitions between even and odd-parity manifolds:
\begin{equation}
L_{\nu,\mu\mu^{\prime}}
=
\sqrt{\gamma_\nu(\omega_{\mu\mu^{\prime}})}
\,
C_{\mu\mu^{\prime}}
\,
|AP\mu\rangle\langle P\mu^{\prime}|, \qquad \mu,\mu'\in\{+,-\},
\end{equation}
with $\omega_{\mu\mu^{\prime}}=E_{AP\mu}-E_{P\mu^{\prime}}$, and the reverse processes given by the Hermitian conjugates at $-\omega_{\mu\mu'}$. Explicit expressions for the coefficients $C_{\mu\mu^{\prime}}$ are given in the Supplemental Material.
\paragraph{First-order polaron correction.}
$S_\nu^{(1)}$ of Eq.~\eqref{eq:Snu1_def_new} induces bath-assisted sideband transitions that simultaneously change the qubit manifold and the cavity occupation,
\begin{equation}
\ket{P\mu',n}\longleftrightarrow\ket{\mathrm{AP}\mu,n\pm1}.
\end{equation}
In the unresolved-Kerr regime the corresponding Bohr frequencies are $\omega_{\mu\mu'}\pm\omega_c$; in the resolved-Kerr regime they acquire the associated number-dependent shifts. The corresponding jump operators are
\begin{equation}
\begin{split}
L_\nu^{(1)}(\omega_{\mu\mu'}+\omega_c)
&=
\sqrt{\gamma_\nu(\omega_{\mu\mu'}+\omega_c)}
\sum_{n\ge0}
\sqrt{n+1}\,\widetilde{C}_{\mu\mu'}
\\
&\qquad
\ket{\mathrm{AP}\mu,n+1}\!\bra{P\mu',n},
\end{split}
\end{equation}
with the reverse processes given by the Hermitian conjugates at negative frequency. 

\subsection{Collision-model derivation}

We now derive the same effective master equation from a collision-model process \cite{CiccarelloCollisionReview,CusumanoCollisionGuide}. See the Supplemental Material for a detailed derivation. 

Within the collision-model perspective, the dressed qubit-cavity system interacts during each short time step $\Delta t$ with fresh ancillas modeling photon and qubit-sector environmental excitations. The one-step map is
\begin{equation}
\rho(t+\Delta t)
=
\Tr_{\rm anc}
\!\left[
U_{\Delta t}
\big(
\rho(t)\otimes \eta
\big)
U_{\Delta t}^\dagger
\right],
\label{eq:coll_map_main_new}
\end{equation}
where $\eta$ is the state of the incoming ancillas and
\begin{equation}
U_{\Delta t}
=
\exp\!\left[
-i\Delta t\,H_{\rm eff}
-i\sqrt{\Delta t}\sum_{r=\gamma,\nu}V_r
\right].
\label{eq:U_coll_main_new}
\end{equation}

Here $H_{\rm eff}$ is the same coherent effective Hamiltonian used in the standard Born--Markov--secular derivation. The $\sqrt{\Delta t}$ scaling is the standard weak-collision scaling leading to a finite dissipative contribution in the continuum limit \cite{AttalPautrat,CiccarelloCollisionReview}.

For each bath label $r\in\{\gamma,\nu\}$, the corresponding collision Hamiltonian is constructed as
\begin{equation}
V_r=
\sum_{\omega>0}
\Big(
S_r(\omega)\otimes b_{r,\omega}^\dagger
+
S_r^\dagger(\omega)\otimes b_{r,\omega}
\Big)
+
S_r(0)\otimes X_r .
\label{eq:Vr_main_new}
\end{equation}
As in the derivation above, the operators $S_r(\omega)$ are the same dressed eigenoperators obtained from the same reference Hamiltonian, namely by neglecting the $H_J$ term of Eq.~\eqref{eq:polaron_exchange} in the dissipative decomposition while retaining it in the coherent dynamics. For the cavity bath, we likewise work within the same unresolved-Kerr approximation.

As shown in the Supplemental Material, assuming vanishing ancilla first moments and expanding Eq.~\eqref{eq:coll_map_main_new} to order $\Delta t$, one obtains in the continuum limit
\begin{equation}
\begin{split}    
\dot\rho
&=
-i[H_{\rm eff}+H_{\rm LS},\rho]
+
\sum_{r,\omega>0}
\Gamma_r^\downarrow(\omega)\,
\mathcal D[S_r(\omega)]\rho\\
&+
\sum_{r,\omega>0}
\Gamma_r^\uparrow(\omega)\,
\mathcal D[S_r^\dagger(\omega)]\rho
+
\sum_r
\Gamma_r^0\,
\mathcal D[S_r(0)]\rho,
\label{eq:GKLS_collision_main_new}
\end{split}
\end{equation}
where the rates are fixed by ancilla second moments.

Equation~\eqref{eq:GKLS_collision_main_new} has the same GKLS structure as the master equation obtained in the standard Born--Markov--secular derivation. Upon identifying the ancilla correlators with the bath spectral rates,
\begin{equation}
\Gamma_r^\downarrow(\omega)=\gamma_r(\omega),\qquad
\Gamma_r^\uparrow(\omega)=\gamma_r(-\omega),\qquad
\Gamma_r^0=\gamma_r(0),
\end{equation}
the collision-model continuum limit reproduces the same effective jump operators derived above, namely the cavity photon channels of Eq.~\eqref{eq:cavity_jumps_new}, the photon-induced dressed antiparallel channels of Eq.~\eqref{eq:photon_qubit_jumps_new}, and the corresponding qubit-sector intra-manifold, cross-manifold, and first-order sideband channels.

This repeated-interaction formulation accommodates both control scenarios discussed in Sec.~\ref{sec:ideal}. In the simultaneous setting, beam-splitter and Kerr terms act within the same coherent step; in the sequential setting, one alternates beam-splitter and Kerr layers with the corresponding dissipative collision steps. 

\section{Average gate fidelity in the weak-dissipation regime}
\label{sec:fidelity_main}
Let us now derive a simplified expression for the average fidelity in the weak-dissipation regime. See Appendix~\ref{app:supp_fidelity} for the full derivation.

For a quantum channel $\mathcal E$ and target unitary $U_g$,
the average gate fidelity over pure states in a $d$-dimensional
Hilbert space is defined as \cite{Nielsen}
\begin{equation}
F_{\mathrm{avg}}
=
\int d\psi\,
\langle\psi|
U_g^\dagger
\mathcal E(|\psi\rangle\langle\psi|)
U_g
|\psi\rangle,
\end{equation}
where $d\psi$ denotes the Haar measure.
In our case, we quantify performance via the average gate fidelity between the implemented noisy channel $\mathcal{E}$ and the ideal unitary $U_g$ over a gate time $t_g$ on the truncated Hilbert space
$\mathcal{H}_d=\mathcal{H}_{\mathrm{qb}}\otimes \mathrm{span}\{\ket{n}\}_{n=0}^{n_{\max}}$, with $d=4(n_{\max}+1)$, see Supplemental Material for a detailed derivation.

Assuming $\gamma_k t_g \ll 1$ for all dissipation rates, 
we expand the evolution to first order in the dissipator.
Evaluating the dissipative correction along the ideal unitary trajectory 
and performing the Haar average over pure states, 
one obtains the general first-order expression

\begin{equation}
F_{\rm avg}
=
1-\frac{t_g}{4(n_{\max}+1)+1}\Big(\Sigma_\gamma+\Sigma_\nu\Big)
+\mathcal O(\gamma^2 t_g^2),
\label{eq:Favg_split_main}
\end{equation}

where we have obtained a simplified expression, since all collapse operators are traceless and we have separated between photon and qubit-sector bath contributions $\Sigma_\gamma=\sum_{k\in\gamma}\Tr(L_k^\dagger L_k)$ and 
$\Sigma_\nu=\sum_{k\in\nu}\Tr(L_k^\dagger L_k)$.

Using the Lindblad operators derived in  Sec.~\eqref{sec:lindblad_main}, we obtain
\begin{equation}
\Sigma_\gamma
=
(n_{\max}+1)(\Sigma_{\gamma,\mathrm{cav}}
+
\Sigma_{\gamma,\mathrm{ap}}),
\label{eq:Sigma_gamma_decomp_main}
\end{equation}
with
\begin{equation}  
\begin{split}
\Sigma_{\gamma,\mathrm{cav}}
&:=
2n_{\max}\big(\kappa_\downarrow+\kappa_\uparrow\big),\\
\Sigma_{\gamma,\mathrm{ap}}
&:=
16\lambda^2\sin^2(2\theta_{\mathrm{ap}})
\big(\gamma_\gamma(\omega_{\mathrm{ap}})+\gamma_\gamma(-\omega_{\mathrm{ap}})\big)
\\
&+32\lambda^2\cos^2(2\theta_{\mathrm{ap}})\gamma_\gamma(0),
\end{split}
\end{equation}
where $\lambda=g/\omega_c$ and $\omega_{\mathrm{ap}}=2\Omega_{\mathrm{ap}}$.

For the qubit-sector bath contribution
\begin{equation}
\Sigma_\nu=(n_{\max}+1)\Big(\Sigma_{\nu, \rm p}+\Sigma_{\nu,\mathrm{ap}}+
\Sigma_{\nu,\times}+\frac{n_{\max}}{2}\Sigma_{\nu,\times}^{(1)}\Big),
\label{eq:Sigma_nu_decomp_main}
\end{equation}
with
\begin{equation}
\begin{split}    
\Sigma_{\nu,\rm p}
&:=(\beta_1+\beta_2)^2
\Big[
\sin^2(2\theta_{\rm p})\big(\gamma_\nu(\omega_{\rm p})\\
&+\gamma_\nu(-\omega_{\rm p})\big)+2\cos^2(2\theta_{\rm p})\gamma_\nu(0)
\Big],\\
\Sigma_{\nu,\mathrm{ap}}
&:=(\beta_1-\beta_2)^2
\Big[
\sin^2(2\theta_{\mathrm{ap}})\big(\gamma_\nu(\omega_{\mathrm{ap}})\\
&+\gamma_\nu(-\omega_{\mathrm{ap}})\big)+2\cos^2(2\theta_{\mathrm{ap}})\gamma_\nu(0)
\Big],
\\
\Sigma_{\nu,\times}
&:=\sum_{\mu,\mu^{\prime}=\pm}
|C_{\mu\mu'}|^2
\big(\gamma_\nu(\omega_{\mu\mu^{\prime}})+\gamma_\nu(-\omega_{\mu\mu^{\prime}})\big),\\
\\
\Sigma_{\nu,\times}^{(1)}
&:=\sum_{\mu,\mu^{\prime},s, s^{\prime}=\pm}
|\widetilde C_{\mu\mu'}|^2
\Big[
\gamma_\nu(s\omega_{\mu\mu'}+s^{\prime}\omega_c)\Big].
\end{split}\label{eq:Fgamma_budget_main}
\end{equation}
Here $\omega_{\rm p}=2\Omega_{\rm p}$ and $\omega_{\mathrm{ap}}=2\Omega_{\mathrm{ap}}$ are the dressed splittings
in the parallel and antiparallel manifolds, and $\omega_{\mu\mu^{\prime}}=E_{\mathrm{AP}\mu}-E_{P\mu^{\prime}}$ are cross-manifold Bohr frequencies.
The overlaps $M^{(i)}_{\mu\mu^{\prime}}=\langle AP\mu|\sigma_x^{(i)}|P\mu^{\prime}\rangle$ are given explicitly in the Supplemental Material.

\section{cQED-CNT spin-qubit design}
\label{sec:cnt_design_recipe}

In this section we discuss a possible physical implementation of the proposed gate in the carbon-nanotube circuit-QED platform developed by C12 Quantum Electronics and by A.~Cottet, T.~Kontos, and M.~M.~Delbecq at the \textit{laboratoire de physique de l'\'{e}cole normale sup\'{e}rieure} (LPENS) \cite{neukelmance2025microsecond}. In this approach, the information is encoded in the states of gate-defined double quantum dots that are hosted in a carbon nanotube. As similar implementations can be made in other semiconducting materials and this is not the core message of the article, we won't expand to much on the reasoning behind the choice of this material for hosting the qubit; more on this topic can be found in \cite{cottet2010spin,penfold2017microwave,delbecq2020,khivrich2020atomic,chen2023long}.

\subsection{The CNT spin qubit}

\begin{figure}
    \centering
    \includegraphics[width=\columnwidth]{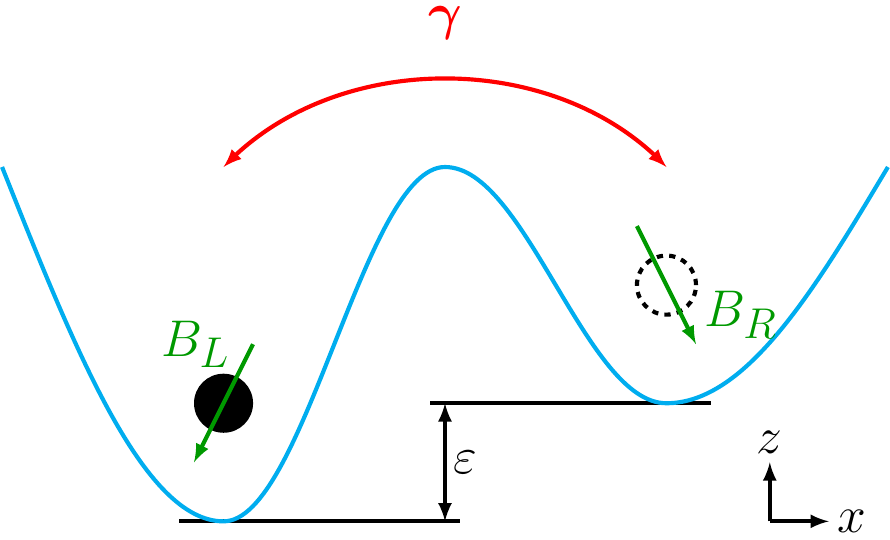}
    \includegraphics[width=\columnwidth]{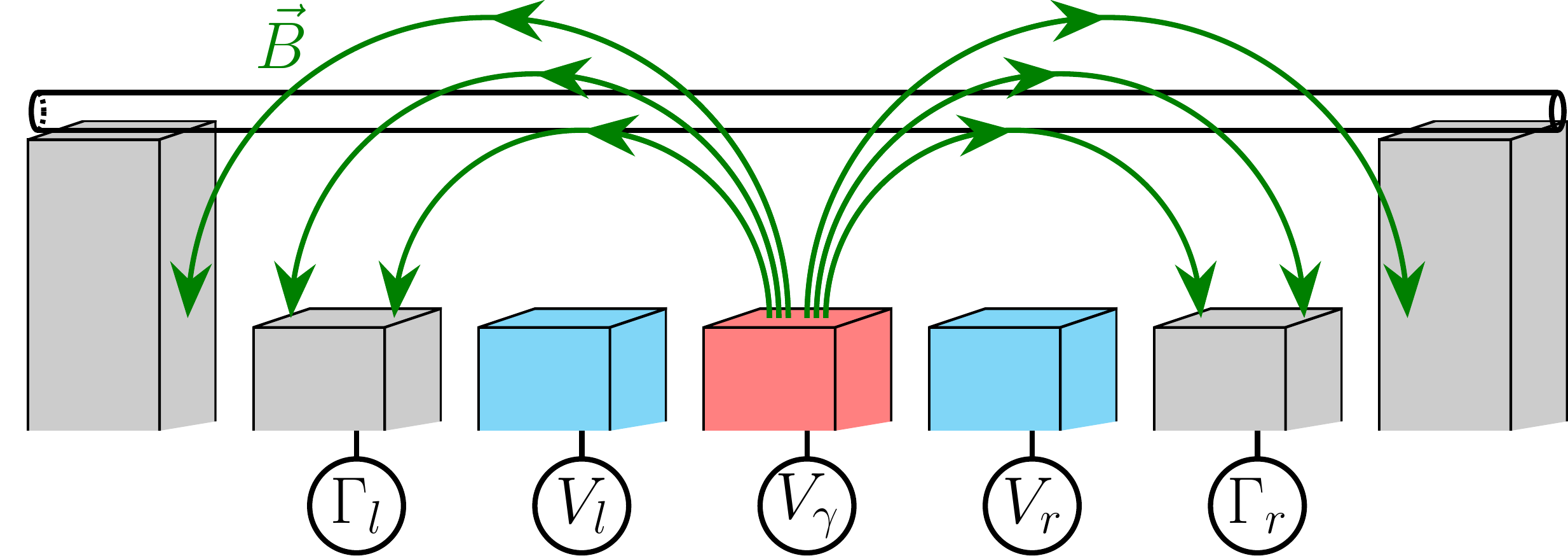}
    \caption{Schematic representation of the CNT spin qubit. \(B\) denotes the magnetic texture, with \(B_{L(R)}\) its projection on the left (right) dot. \(\varepsilon\) is the DQD bias energy and \(\gamma\) the interdot tunneling amplitude. \(\Gamma_j\) with \(j\in\{l,r\}\) denote the voltages applied to the outer gates and control the DQD-contact tunneling rates. \(V_k\) with \(k\in\{l,\gamma,r\}\) denote the voltages applied to the central gates, which control the DQD bias and tunneling energy.}
    \label{fig1}
\end{figure}

We first briefly review the CNT spin-qubit architecture developed at LPENS and C12. The basic ingredients are shown in Fig.~\ref{fig1}. A suspended single-wall CNT is stapled \cite{cubaynes2019highly} between source and drain contacts above five electrostatic gates, which define a single-electron DQD. The orbital degree of freedom of this DQD forms a charge qubit. A magnetic texture, generated either by a micromagnet \cite{legrand2023optimal} or by ferromagnetic contacts \cite{cottet2010spin}, induces an artificial spin-orbit interaction together with a Zeeman splitting, thereby enabling spin manipulation. This defines the CNT spin qubit.

The spin-qubit Hamiltonian can be written as
\begin{equation}
    H_q=\frac{\varepsilon}{2}\tau_z+\gamma\tau_x+\frac{\alpha_s}{2}\sigma_z+\frac{\alpha_{as}}{2}\sigma_x\tau_z,
    \label{hq}
\end{equation}
where \(\varepsilon\) is the DQD bias, \(\gamma\) the interdot tunneling amplitude, \(\alpha_s=B_L+B_R\) the symmetric component of the magnetic potential, and \(\alpha_{as}=B_L-B_R\) its antisymmetric component. Here \(\sigma_k\) (\(k\in\{x,y,z\}\)) denote spin Pauli operators, while \(\tau_k\) act on the orbital degree of freedom, i.e.\ on the electron position in the DQD. Diagonalizing \(H_q\) yields the qubit eigenbasis; further details can be found in Ref.~\cite{benito2019optimized}.

The resonator mediating the two-qubit interaction is typically coupled to one of the plunger gates, \(V_l\) or \(V_r\) in Fig.~\ref{fig1}. In the DQD basis, the electric-dipole coupling reads
\begin{equation}
    H_d=g\tau_z(a+a^\dagger).
\end{equation}
In the qubit representation this becomes \cite{benito2019optimized}
\begin{equation}
    \bar{H}_d=(a+a^\dagger)\left(g_\sigma\sigma_x\tau_z+g_\tau\tau_z+\lambda_\sigma\sigma_z+\lambda_\tau\tau_z\right),
\end{equation}
where the coefficients \(g_k\) and \(\lambda_k\) depend on the experimentally controlled parameters \(\varepsilon\) and \(\gamma\). When the DQD is far detuned, the electron is mainly located on one dot and the electric dipole is suppressed. In that regime the spin is isolated and both couplings are vanishing. At $\varepsilon=0$, the coupling is maximal and fully transverse, meaning that $\lambda_k=0$. This illustrates how gate voltages can be used to tune both the strength and the geometry of the qubit-resonator coupling, or even suppress it altogether.

\subsection{CNT-CQED hybrid beam splitter and Kerr}

The implementation of our proposed scheme for a qubit-boson hybrid beam-splitter gate rely on two qubits being transversely coupled together and longitudinally coupled to the same resonator. In addition to that, we require a tunable Kerr anharmonicity in the resonator.

Let us first discuss the transverse coupling between the two qubits. A natural route is to define two qubits on the same CNT by introducing additional control gates under the nanotube, thereby forming an array of quantum dots coupled through an exchange-type interaction \cite{loss1998quantum}. This type of multi-dot control has recently been demonstrated at LPENS in the group of T.~Kontos and M.~M.~Delbecq \cite{craquelin2026electrical}, using a sixteen-gate CNT device. In our proposal, nine tunable gates would be sufficient, as sketched in Fig.~\ref{fig:cnt_design}: two source and drain gates, six gates defining the two qubits, while a central gate controls their separation and interaction.

When the tunneling between the two DQDs is sufficiently weak, the dominant inter-qubit coupling is capacitive and takes the form \(\tau_z^{(1)}\tau_z^{(2)}\). Starting from Eq.~\eqref{hq}, the resulting two-qubit Hamiltonian is
\begin{equation}
    H_{2q}=\sum_{k\in\{1,2\}}H_q^{(k)}+J\tau_z^{(1)}\tau_z^{(2)}.
\end{equation}
At \(\varepsilon_k=0\), and after projection onto the qubit subspace, this reduces to
\begin{equation}
    \bar{H}_{2q}=\sum_{k\in\{1,2\}}\bar{H}_q^{(k)}+J\sigma_x^{(1)}\sigma_x^{(2)}.
\end{equation}
Here we restrict to the low-energy qubit sector, i.e.\ to states satisfying \(\langle\tau_z\rangle=-1\). By detuning the dots so that the electrons occupy the outermost sites, this interaction can be strongly suppressed. The suppression comes both from the increased spatial separation between the electrons and from the fact that the qubits approach a pure-spin regime in this configuration.

We now turn to the spin-photon coupling. In the present proposal, the qubit-resonator coupling must be longitudinal during the beam-splitter stage, while it should be strongly suppressed during the Kerr stage. If the resonator is connected to a plunger gate, then at \(\varepsilon_k=0\) the coupling is purely transverse. To obtain a longitudinal coupling at \(\varepsilon_k=0\), the resonator electric field must instead couple to an electronic transition in the DQD, so that in the DQD basis the interaction is proportional to \(\tau_x\) rather than \(\tau_z\).
From Eq.~(15) of Ref.~\cite{cottet2015electron}, the relevant matrix elements are of the form
\begin{equation}
    g_{nm}=-e\int V_\perp(x)\phi_n(x)\phi_m^\ast(x)\,d^3x,
\end{equation}
where \(\phi_n(x)\) denotes the orbital wavefunction labeled by \(n\). This suggests coupling the resonator to the interdot barrier gate rather than to a plunger gate: in that case, the resonator modulates interdot tunneling and generates a \(\tau_x^{(k)}\) coupling in the DQD basis, which maps to a longitudinal \(\sigma_z^{(k)}\) coupling in the spin-qubit basis at \(\varepsilon_k=0\).

A further requirement of the effective model is that the two longitudinal couplings have opposite signs. This can be engineered by coupling the two qubits to opposite ends of the resonator, so that the resonator voltage fluctuations seen by the two DQDs have opposite phase.

A further requirement is that the longitudinal spin-photon coupling be tunable, so that it can be active during the beam-splitter stage and strongly suppressed during the Kerr stage. This can be achieved by adiabatically transferring the electrons to opposite ends of the device. In that regime, both the capacitive coupling and the spin-charge hybridization are suppressed while the encoded quantum state is preserved.

We now turn to the last ingredient of the proposal, namely a tunable Kerr nonlinearity. A simple route is to embed a SQUID into the resonator, following standard cQED implementations of nonlinear microwave modes based on Josephson elements \cite{Bourassa2012}. 
A term of the form
\begin{equation}
H_K=\chi\,a^{\dagger 2}a^2
\end{equation}
can then be incorporated into a microwave resonator by embedding a nonlinear inductive element into the cavity. Equivalently, one may write the nonlinearity as a term proportional to \(n^2=(a^\dagger a)^2\), which differs from \(a^{\dagger 2}a^2=n(n-1)\) only by a linear-in-\(n\) renormalization of the mode frequency. A natural route is to terminate or interrupt the resonator with a Josephson element, such as a single Josephson junction, a SQUID, or a SNAIL. In that case, the cosine Josephson potential generates an effective quartic contribution to the cavity Hamiltonian, which yields a self-Kerr interaction after quantization. In particular, replacing a single junction by a SQUID makes the Kerr coefficient flux tunable \textit{in situ}, which is especially appealing when one wishes to work in a weak but controllable nonlinear regime \cite{Bourassa2012}.

For the parameter range relevant here, namely a resonator in the GHz range with a weak but tunable self-Kerr nonlinearity in the MHz range, this regime is experimentally realistic. A recent experiment with a SNAIL-terminated superconducting resonator reported a flux-tunable Kerr coefficient spanning approximately \(-5\) to \(+6~\mathrm{MHz}\), and demonstrated operation at \(K/2\pi \simeq 5.21~\mathrm{MHz}\) for cat-state generation \cite{Lu2023Cat}. This shows that a moderate, tunable Kerr nonlinearity of the size required for our implementation is accessible with present-day Josephson-resonator technology. An alternative, potentially attractive in hybrid spin-based platforms, is to use a nonlinear kinetic-inductance element based on superconducting nanowires rather than Josephson junctions. Such devices are more compatible with operation in large magnetic fields, but current microwave demonstrations report substantially smaller self-Kerr shifts: for example, Ref.~\cite{Joshi2022} reports \(K/2\pi \simeq 123.5~\mathrm{kHz}\) for a nanowire resonator with resonance frequency \(f_0 \simeq 6.3~\mathrm{GHz}\). At the same time, nonlinear nanowire resonators have been operated in magnetic fields up to \(2~\mathrm{T}\) \cite{Khalifa2023}. Therefore, for the present target values, a SQUID- or SNAIL-based design appears to be the most direct implementation route, while nanowire-based nonlinear inductors remain an interesting field-compatible alternative.
\subsection{CNT cQED operating point}\label{sec:cnt_operating_point}
To make the proposed CNT cQED design more concrete, we now identify a representative operating point within the experimentally motivated parameter window summarized in Table~\ref{tab:params}, guided by the realistic device and spectroscopy scales reported in Ref.~\cite{neukelmance2025microsecond}. The goal of this subsection is to verify that the sequential protocol can be placed in a realistic regime and that the main approximations underlying the effective description remain self-consistent there.

In this realization, the qubit-sector environment introduced in Sec.~\ref{sec:lindblad_main} is naturally associated with carbon-nanotube vibrational modes, while the cavity bath remains the photon environment of the resonator. As a representative operating point, we take
\(\omega_1/2\pi=4\,\mathrm{GHz}\),
\(\omega_2/2\pi=7\,\mathrm{GHz}\),
\(g/2\pi=J/2\pi=170\,\mathrm{MHz}\)
and \(\chi/2\pi=5.1\,\mathrm{MHz}\). The cavity frequency is chosen to satisfy the resonance condition with the odd-parity dressed splitting, yielding \(\omega_c/2\pi\simeq 3\,\mathrm{GHz}\). With this choice, the polaron parameter remains moderately small, \(\lambda=g/\omega_c\simeq 5.67\times10^{-2}\), the direct exchange remains perturbative with respect to the qubit detuning, \(J/|\Delta|\simeq 5.7\times10^{-3}\) and the Kerr scale stays weak compared with the bare cavity frequency, \(\chi/\omega_c\simeq 1.7\times10^{-3}\).

At the same operating point, the effective beam-splitter coupling becomes
\begin{equation}
g_{\mathrm{eff}}=\frac{4Jg}{\omega_c},
\qquad
g_{\mathrm{eff}}/2\pi \simeq 3.85~\mathrm{MHz},
\end{equation}
which corresponds to the reference beam-splitter time
\begin{equation}
t_{\mathrm{BS}}=\frac{\pi}{2g_{\mathrm{eff}}}\simeq 64.9~\mathrm{ns}.
\end{equation}
For Kerr-stage durations in the range \(t_{\mathrm K}\sim 10\text{--}30~\mathrm{ns}\), the full sequential gate time therefore lies in the interval
\begin{equation}
t_g=t_{\mathrm{BS}}+t_{\mathrm K}\simeq 74.9\text{--}94.9~\mathrm{ns}.
\end{equation}

A second consistency check concerns dissipation. For the cavity, taking \(Q_c=3\times10^4\) gives
\begin{equation}
\kappa=\frac{\omega_c}{Q_c},
\qquad
\kappa/2\pi \simeq 0.10~\mathrm{MHz},
\end{equation}
while the phonon-induced effective rates are chosen in the range \(10^{1}\text{--}10^{2}\,\mathrm{s}^{-1}\), consistent with MHz-scale vibrational frequencies and \(Q_\nu\sim 10^{5}\). As a result, the phonon-sector contributions satisfy \(\gamma_\nu t_g \ll 1\) throughout the relevant operating window. The cavity channel is less asymptotically separated, but still remains perturbative over the gate duration, so the weak-dissipation treatment may be regarded as controlled, even if not parametrically extreme. This is precisely why, in the next subsection, we benchmark the analytical formulas directly against the full numerical Lindblad evolution.

At low temperature, \(T=10~\mathrm{mK}\), the upward thermal rates associated with GHz-scale transitions are exponentially suppressed and may be neglected to a very good approximation, while the zero-frequency rates remain finite. At the same operating point, the first-order analytical fidelity formula of Eq.~\eqref{eq:Favg_split_main} gives
\begin{equation}
1-F_{\mathrm{avg}} \approx 1.46\times10^{-2},
\qquad
F_{\mathrm{avg}} \approx 0.98537,
\end{equation}
with normalized error contributions
\begin{equation}
\begin{split}
&\frac{E_\gamma}{\sum_j E_j} \approx 99.997\%,\quad
\frac{E_\nu^{(\alpha)}}{\sum_j E_j} \approx 1.91\times10^{-3}\%,\\
&\frac{E_\nu^{(\beta)}}{\sum_j E_j} \approx 1.09\times10^{-3}\%,\quad
\frac{E_\nu^{(1)}}{\sum_j E_j} \approx 4\times10^{-6}\%.
\end{split}
\end{equation}
Thus, in the representative CNT cQED regime considered here, the infidelity is overwhelmingly dominated by the cavity-photon channel, while all phonon-induced contributions remain strongly subleading. Accordingly, improved performance is expected primarily from reducing cavity loss or shortening the total gate duration. In practice, this can be achieved by increasing the cavity quality factor \(Q_c\), or by accelerating the coherent exchange dynamics through a larger effective beam-splitter coupling \(g_{\mathrm{eff}}\), for instance via larger \(g\) and/or \(J\), provided the hierarchy assumptions underlying the effective model remain satisfied. Moreover, the cavity truncation also has a quantitative impact on the reported average gate fidelity. Here we assumed \(n_{\max}=3\), so the benchmark is performed over the full truncated qubit-cavity Hilbert space of dimension \(4(n_{\max}+1)\). This provides a conservative estimate, since cavity loss contributes more strongly as the accessible bosonic manifold is enlarged. In practice, if the protocol is intended to operate predominantly within a lower-excitation sector, restricting the benchmark to that sector would lead to a less punitive fidelity measure.

Altogether, this operating point places the protocol in a realistic CNT cQED regime where the effective hierarchy underlying the gate construction is preserved, the Kerr stage remains a secondary nonlinear scale and dissipation is weak enough to admit an analytical treatment while still being strong enough to require explicit numerical validation.

\begin{table}[t]
\centering
\caption{Representative CNT cQED parameter ranges}
\label{tab:params}
\begin{tabular}{lc}
\toprule
Parameter & Value \\
\midrule
$T$ & $10~\mathrm{mK}$ \\
$\omega_1 \sim\omega_2\sim \omega_c$ & $
~\mathrm{GHz}$ \\
$g$ & $20-200~\mathrm{MHz}$ \\
$J$ & $10^{-1}g$ \\
$\chi$ & $10^{-2}-10^{-1} \ g$ \\
\midrule
$Q_{\gamma}$ & $ 10^{2}-10^{4}$\\ 
$Q_{\nu}$ & $ 10^{5}-10^{6}$\\ 
\bottomrule
\end{tabular}
\end{table}

%

\section{Numerical simulations}
\label{sec:simulations}

To benchmark the effective open-system description, we simulate the Lindblad master equation on the truncated Hilbert space
\begin{equation}
\mathcal H_d=\mathcal H_{\mathrm{qb}}\otimes \mathrm{span}\{\ket{n}\}_{n=0}^{n_{\max}},
\end{equation}
using the Python-based QuTiP library \cite{Johansson2012QuTiP,Johansson2013QuTiP}. The numerical evolution is generated by the effective Hamiltonian and the collapse operators derived in Sec.~\ref{sec:lindblad_main}. Unless otherwise stated, we use the same operating-point parameter set of Section ~\ref{sec:cnt_operating_point}. Specifically, the cavity truncation \(n_{\max}\) is chosen large enough that the occupation of the cutoff level remains negligible throughout the gate evolution. In the low-excitation two-qubit regime considered here, a suitable cutoff is largely dictated by the block structure of the effective beam-splitter interaction. Since the hybrid exchange couples only neighboring sectors, \(\ket{AP-,n}\leftrightarrow\ket{AP+,n+1}\), an initial state supported on low Fock numbers spreads only locally in number space. It is therefore sufficient, for the benchmarks reported here, to choose the truncation a small number of levels above the highest initially occupied sector. A more systematic truncation analysis would become increasingly important in lattice extensions or in regimes with larger bosonic occupation, where one would instead need to verify the convergence of representative cavity, qubit, and reduced-state diagnostics as the cutoff is increased.

The following simulations serve two complementary purposes. First, they validate the effective noiseless and noisy gate dynamics and its weak-dissipation analytical description. Second, they access to the reduced cavity-state structure generated by the hybrid qubit-cavity evolution.
\subsection{Gate dynamics and fidelity benchmark}
\begin{figure*}[t]
    \centering
    \begin{subfigure}[t]{0.32\textwidth}
        \centering
        \includegraphics[width=\textwidth]{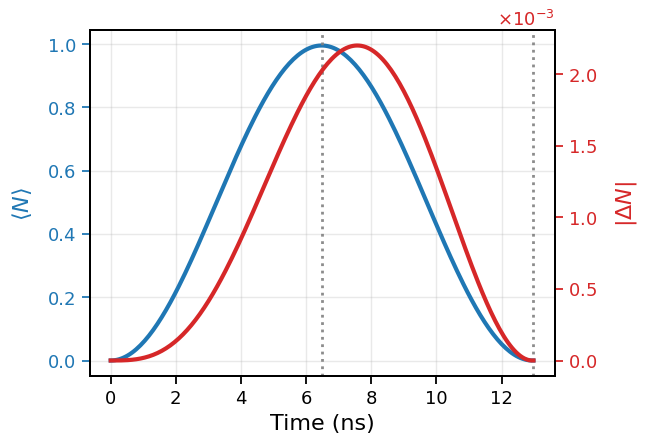}
    \end{subfigure}
    \hfill
    \begin{subfigure}[t]{0.32\textwidth}
        \centering
        \includegraphics[width=\textwidth]{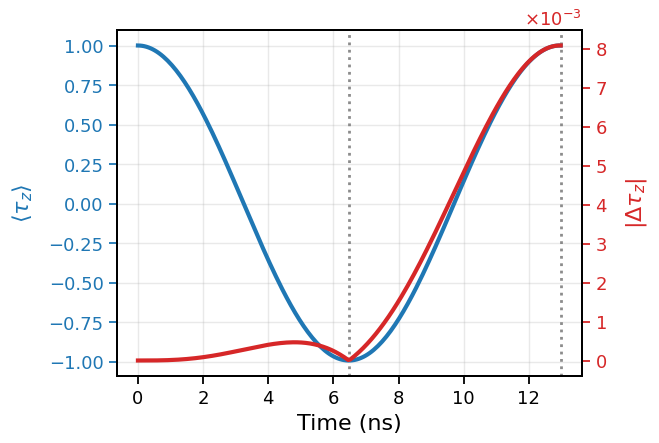}
    \end{subfigure}
    \hfill
    \begin{subfigure}[t]{0.32\textwidth}
        \centering
        \includegraphics[width=\textwidth]{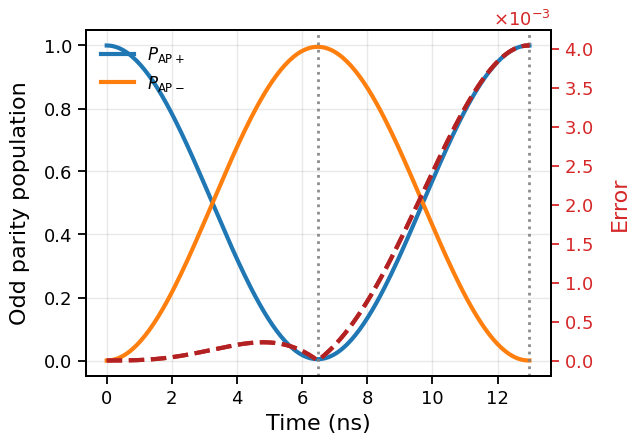}
    \end{subfigure}

    \vspace{0.2em}

    \makebox[\textwidth][c]{%
        \begin{subfigure}[t]{0.32\textwidth}
            \centering
            \includegraphics[width=\textwidth]{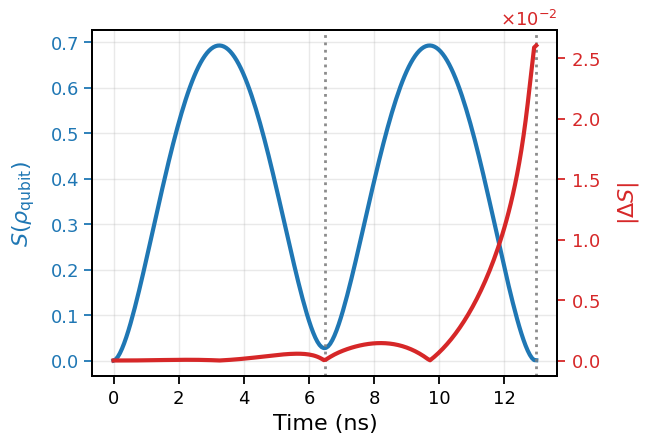}
        \end{subfigure}
        \hspace{0.04\textwidth}
        \begin{subfigure}[t]{0.32\textwidth}
            \centering
            \includegraphics[width=\textwidth]{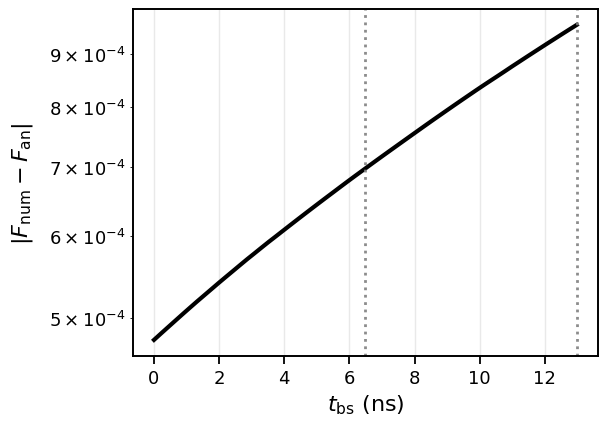}
        \end{subfigure}
    }
   \caption{
Numerical benchmark of the effective open-system gate dynamics for the representative CNT cQED operating point discussed in the text, using a cavity truncation \(n_{\max}=3\) and initial state \(\ket{AP+,0}\). The top row shows pure beam-splitter evolution over two reference exchange times \(t_{\mathrm{ref}}=\pi/(2g_{\mathrm{eff}})\). In each panel, the left axis reports the ideal trajectory, while the right axis shows the absolute deviation between noisy and ideal evolution. From left to right, the panels display the cavity occupation \(\langle N\rangle(t)\), the dressed qubit polarization \(\langle \tau_z\rangle(t)\), and the odd-parity dressed-state populations \(P_{\mathrm{AP}+}\) and \(P_{\mathrm{AP}-}\). The bottom-left panel shows the reduced qubit entropy, again together with the corresponding noisy-ideal deviation. 
The bottom-right panel reports the absolute residual \(|F_{\mathrm{num}}-F_{\mathrm{an}}|\) between the numerical gate fidelity and the analytical weak-dissipation prediction for the sequential protocol, as a function of the beam-splitter duration \(t_{\mathrm{bs}}\) at fixed \(t_k=30\,\mathrm{ns}\). The residuals remain at the \(10^{-4}\) level, indicating good agreement between the analytical and numerical fidelities in the operating regime considered here. 
}
    \label{fig:dynamics_main}
\end{figure*}

Figure~\ref{fig:dynamics_main} shows the gate dynamics and benchmarks the effective open-system description against the corresponding ideal evolution. To isolate the exchange dynamics generated by the hybrid beam-splitter interaction, the top-row panels show pure beam-splitter evolution with \(t_K=0\), starting from the initial state \(\ket{AP+,0}\) in the odd-parity manifold. The cavity occupation \(\langle N\rangle(t)\) and the dressed polarization \(\langle \tau_z\rangle(t)\) display the coherent back-and-forth transfer of excitation between the dressed qubit sector and the cavity mode, while the odd-parity populations \(P_{\mathrm{AP}+}\) and \(P_{\mathrm{AP}-}\) resolve the same process directly within the active qubit manifold. In each case, the ideal trajectory is shown together with the absolute deviation induced by dissipation, making clear that the exchange dynamics remains visible while being gradually degraded by the cavity-photon and phonon environments. Over the time window shown, these deviations remain at the \(10^{-2}\text{--}10^{-3}\) level relative to the corresponding ideal observables, confirming that dissipation acts as a perturbative correction to the coherent exchange dynamics. The reduced qubit entropy provides a complementary diagnostic of the hybrid dynamics by monitoring the loss of local qubit purity generated by qubit-cavity correlations and, in the noisy case, by the action of the environment.

The same figure also summarizes the gate-performance benchmark for the sequential protocol. The fidelity residual \(|F_{\mathrm{num}}-F_{\mathrm{an}}|\), shown as a function of the beam-splitter duration \(t_{\mathrm{BS}}\) at fixed Kerr-stage duration, provides a direct measure of the accuracy of the analytical weak-dissipation treatment. The residuals remain small throughout the relevant operating window, indicating that the analytical fidelity formula captures the numerical Lindblad dynamics well in the representative CNT cQED regime considered here. We stress that these benchmarks are performed with cavity truncation \(n_{\max}=3\), which is sufficient for the low-excitation dynamics explored from the initial state considered here. Increasing \(n_{\max}\) makes the full-space fidelity benchmark more stringent, since dissipation acts over a larger truncated bosonic manifold.

\subsection{Reduced cavity-state structure under hybrid exchange}

To further characterize the cavity-state structure generated by the effective gate dynamics, we monitor the reduced cavity mode through the Gaussian-closure deviation and the Wigner function. For the first, we consider
\begin{equation}
\Delta_G(t)=
\langle a^\dagger a^\dagger a a\rangle
-2\langle a^\dagger a\rangle^2
-|\langle a a\rangle|^2,
\end{equation}
evaluated on the reduced cavity state. For a single-mode Gaussian state, fourth-order moments are fixed by second-order moments, so \(\Delta_G=0\) is consistent with Gaussian closure. A nonzero value therefore signals a departure from this closure relation, providing a simple indicator of nontrivial reduced cavity-state structure.

\begin{figure*}[t]
    \centering
    \begin{minipage}[c]{0.34\textwidth}
        \centering
        \includegraphics[width=\textwidth]{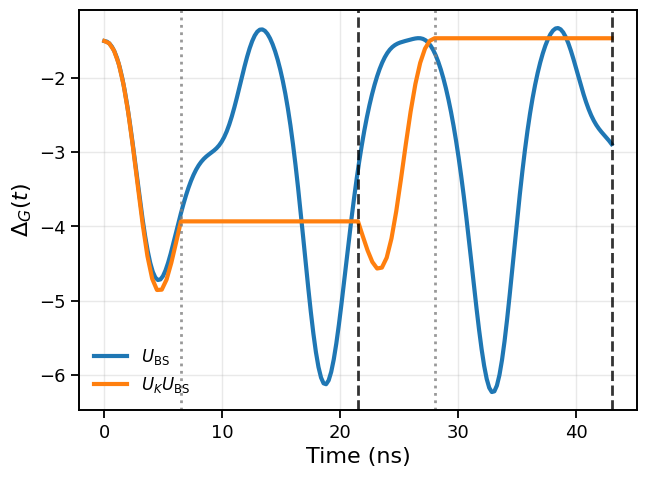}
    \end{minipage}
    \hfill
    \begin{minipage}[c]{0.62\textwidth}
        \centering
        \includegraphics[width=\textwidth]{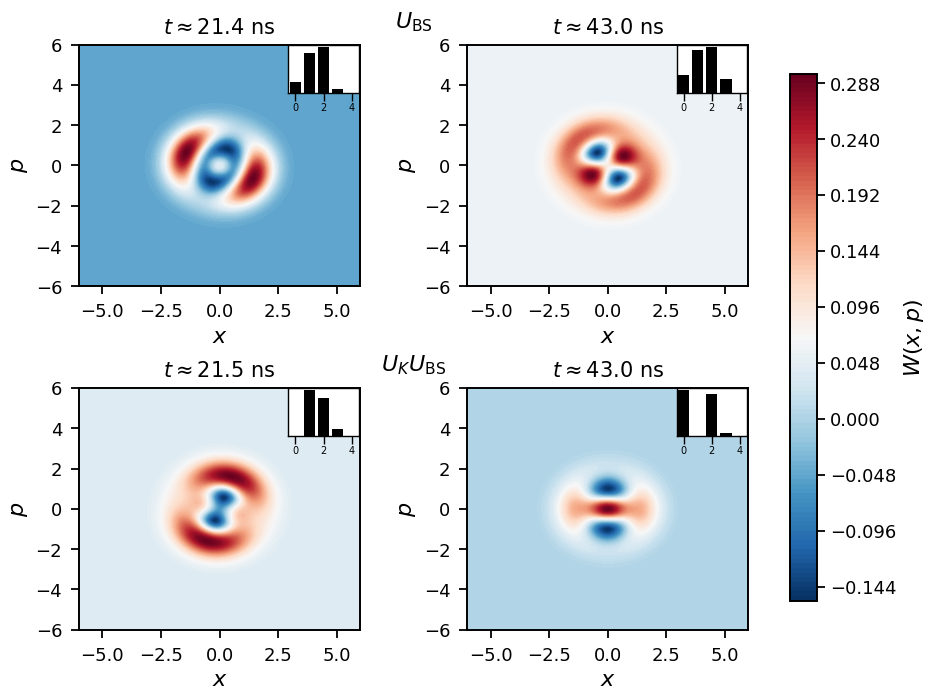}
    \end{minipage}
    \caption{Reduced cavity-state structure generated by the effective gate dynamics, for the initial state \(\ket{AP+}\otimes(\ket{0}+\ket{2})/\sqrt{2}\), cavity truncation \(n_{\max}=8\), and representative Kerr-stage duration \(t_k=15\,\mathrm{ns}\). The left panel shows the Gaussian-closure deviation \(\Delta_G(t)\) of the reduced cavity state, comparing pure beam-splitter evolution \(U_{\mathrm{BS}}\) with the repeated sequential protocol \(U_KU_{\mathrm{BS}}\). Vertical dotted and dashed lines mark, respectively, the internal beam-splitter-to-Kerr switch and the end of each full sequential cycle. The right panels display representative reduced-cavity Wigner snapshots after one and two full cycles for pure beam-splitter evolution (top) and for the sequential protocol (bottom). These results show that the dominant departure from Gaussian closure is already generated by the hybrid beam-splitter dynamics, while the Kerr stage primarily reshapes the resulting phase-space structure through number-dependent phase accumulation.
}
    \label{fig:cavity_structure_main}
\end{figure*}

Representative results are shown in Fig.~\ref{fig:cavity_structure_main} for the initial state \(\ket{AP+}\otimes(\ket{0}+\ket{2})/\sqrt{2}\). We compare pure beam-splitter evolution with the repeated sequential protocol \(U_KU_{\mathrm{BS}}\), in which each cycle consists of a beam-splitter stage followed by a Kerr stage. The time evolution of \(\Delta_G(t)\) shows that the dominant departures from Gaussian closure already arise under the hybrid beam-splitter dynamics alone. This indicates that the exchange stage is the primary mechanism responsible for generating the observed cavity-state structure, while the Kerr stage acts mainly by dressing that state through number-dependent phase accumulation.

This picture is consistent with the representative Wigner snapshots in Fig.~\ref{fig:cavity_structure_main}. In the regime explored here, pure beam-splitter evolution already produces structured reduced cavity states, including negative Wigner regions for suitable initial conditions. The repeated \(U_KU_{\mathrm{BS}}\) sequence does not appear to be the primary source of this nonclassical phase-space structure, but it visibly modifies its morphology over successive cycles, changing the orientation and relative weight of the interference features without qualitatively altering the low-lying Fock-sector support shown in the insets.

\section{Conclusion}\label{sec:conclusion}

In this work, we introduced a hybrid qubit-boson beam-splitter gate acting between an exchange-dressed two-level qubit subsystem and a microwave cavity mode in the presence of Kerr nonlinearity. Starting from a general effective cQED model, we derived the corresponding effective qubit-cavity exchange interaction and discussed its composition with Kerr evolution in sequential control regimes (Sec.~\ref{sec:ideal}).

We then developed an open-system description including photon- and qubit-sector-bath-induced dissipation. The same effective master equation was obtained in two complementary ways, namely from a standard Born--Markov--Davies treatment and from a repeated-interaction collision-model construction \cite{BreuerPetruccione,Davies1974,CiccarelloCollisionReview,CusumanoCollisionGuide}. In the weak-dissipation regime, this allowed us to derive a compact analytical expression for the average gate fidelity (Secs.~\ref{sec:lindblad_main} and \ref{sec:fidelity_main}).

We also identified a concrete carbon-nanotube circuit-QED route to implementing the proposed gate (Sec.~\ref{sec:cnt_design_recipe}). In this setting, two gate-defined CNT spin qubits can be formed on the same nanotube and coupled through an exchange-like interaction, while a common microwave resonator provides the bosonic mode required for the hybrid gate. By appropriately engineering the gate layout, the resonator can be coupled to electronic transitions in each DQD in such a way as to realize an effective longitudinal spin-photon interaction in the qubit basis. Moreover, placing the two qubits at opposite ends of the resonator allows one to control the relative sign of this coupling. Combined with a tunable Kerr nonlinearity, achievable in standard Josephson-resonator architectures such as SQUID- or SNAIL-based designs, this yields a realistic hardware route toward both the beam-splitter and Kerr stages of the protocol. The corresponding cQED implementation and CNT spin-qubit architecture are illustrated in Figs.~\ref{fig:cnt_design} and \ref{fig1}. In this realization, the cavity bath is naturally associated with photon loss, while the qubit-sector bath is microscopically identified with CNT phonons.

Using a representative set of experimentally realistic operating parameters for the CNT cQED design, we then studied the gate dynamics quantitatively (Secs.~\ref{sec:cnt_design_recipe} and \ref{sec:simulations}).
In particular, the numerical simulations reported in Fig.~\ref{fig:dynamics_main} show that the noisy dynamics remains relatively close to the ideal one over the operating regime considered here, with deviations of order \(10^{-2}\)–\(10^{-3}\). The dynamical plots clearly show how the hybrid gate drives a coherent exchange between the cavity mode and the exchange-dressed odd-parity sector. As illustrated by the cavity-state analysis of Fig.~\ref{fig:cavity_structure_main}, this already leads at the beam-splitter stage to nonclassical structure in the reduced cavity state, which is then mainly reshaped by the Kerr stage through phases that depend quadratically on the photon number.

At the same operating point, we also analyzed the gate fidelity and its parameter dependence (Secs. ~\ref{sec:cnt_design_recipe}, and \ref{sec:simulations}). We found that the dominant contribution to the infidelity comes from cavity dissipation, since the gate is not operated in a dispersive regime and therefore pays the full price of cavity loss. More precisely, for the representative operating point identified in Sec.~\ref{sec:cnt_design_recipe}, cavity dissipation accounts for $\approx 99.99 \%$ of the gate error, while the biggest phonon contribution accounts for $\approx 10^{-3} \%$. Accordingly, the performance of the protocol depends primarily on two ingredients: the cavity quality factor and the total gate time, given by the sum of the beam-splitter and Kerr durations. The beam-splitter time is controlled by the effective coupling \(g_{\mathrm{eff}}\propto gJ/\omega_c\), and can therefore be shortened by increasing \(g\) and \(J\) while respecting the hierarchy assumptions required by the effective derivation. The Kerr time is instead set by the nonlinear phase accumulation and can be tuned experimentally via flux control in a SQUID- or SNAIL-based nonlinear resonator. For the representative parameter set considered here, with \(g\sim J\sim 170~\mathrm{MHz}\), beam-splitter time $\approx 65 ~\mathrm{ns}$, cavity quality factor \(Q\sim 3\times10^4\), and Kerr time \(t_K=15~\mathrm{ns}\), we obtain a fidelity of $\approx 98.5 \%$. Improving either the cavity quality factor or the overall gate speed should therefore provide a direct route toward higher fidelities.

Finally, Fig.~\ref{fig:dynamics_main} also provides a direct benchmark of the analytical fidelity formula against full QuTiP simulations. In the weak-dissipation regime considered here, corresponding in particular to the truncation \(n_{\max}=3\), we find good agreement between the numerical and analytical fidelities, with discrepancies of order \(\sim 10^{-4}\). We stress that the fidelity also depends on the cavity truncation, since larger accessible bosonic sectors generally increase the effective spread of the dynamics and the sensitivity to dissipation, thus a careful truncation has to be set in place when performing simulations, as illustrated for instance in the Supplemental Material.

Beyond its direct gate interpretation, the present hybrid primitive also points toward several broader directions. First, it provides a natural building block for locally updated hybrid many-body dynamics, including quantum-cellular-automaton and lattice-gauge-inspired constructions \cite{SellapillayArrighiDiMolfettaQCAQED}, in which a persistent bosonic mode mediates matter-field exchange and carries local field information across successive updates. From this viewpoint, the explicit qubit-cavity implementation considered here realizes the same local mechanism, while the Kerr term supplies the nonlinear number-dependent phases required by the update rule to preserve gauge invariance. If the cavity is not reset between successive updates, the persistence of the bosonic mode can also naturally act as a source of memory.

A related perspective concerns nonclassical and potentially non-Gaussian many-body dynamics. In the present architecture, the emergence of nontrivial bosonic structure might come from the combined action of the beam-splitter and Kerr stages. Our analysis indicates that the beam-splitter stage already generates nonclassical cavity states, while the Kerr stage further reshapes and rephases the resulting bosonic structure. At the same time, from the bosonic point of view Kerr remains the explicit nonlinear ingredient of the dynamics. The hybrid beam-splitter gate acts nontrivially only within the exchange-dressed odd-parity sector of the two-qubit manifold, thereby suggesting a structured parity-preserving dynamics on the qubit side reminiscent of matchgate- or fermionic-linear-optics-type circuits \cite{JozsaMiyakeMatchgates,BrodMatchgates}. From this viewpoint, the zero-Kerr limit provides a useful structured baseline combining a parity-preserving qubit sector with a linear bosonic mediator, while Kerr may be viewed as the ingredient that drives the dynamics away from that simpler regime \cite{WeedbrookGaussianQI,OlsenCorneyKerrNonGaussian,YanagimotoKerrCubic}. In this sense, recent work on dissipative non-Gaussian many-body dynamics motivates further investigation of many-body protocols based on this primitive gate, particularly with a view toward understanding possible growth in classical simulation complexity \cite{SpagnoloNonlinearBosonSampling,DiasKoenigNonGaussianBosonicCircuits,GonzalezGarciaNonGaussianDissipation}.

A second direction concerns open-system simulation and stroboscopic architectures. The collision-model reformulation is useful as an operational language for organizing the dynamics into coherent and dissipative layers. In this language, the present architecture admits a stroboscopic interpretation in which beam-splitter and Kerr windows are interleaved with effective dissipative slices associated with cavity and phonon environments. This is naturally aligned with repeated-interaction simulation strategies \cite{CiccarelloCollisionReview,CusumanoCollisionGuide}. Although the CNT cQED device is not microscopically a stream-of-ancillas platform, the collision-model description still provides a useful conceptual bridge to ancilla-based views of open-system dynamics.

A third direction concerns reservoir-style quantum information processing. Quantum reservoir computing is attractive because it exploits a fixed noisy dynamical system as a high-dimensional nonlinear reservoir and trains only a classical readout layer, making it naturally suited to analog and near-term hardware \cite{GhoshQRP,AraizaBravoQRC,SanniaDissipativeQRC}. In such settings, dissipation is not only a limitation, but it can also provide the fading-memory mechanism required for temporal information processing \cite{SanniaDissipativeQRC}. From this viewpoint, collision models offer a useful conceptual language for repeated input injection, controlled forgetting, and tunable memory, while noisy QCA \cite{AvalleSerafiniNoisyQCA} suggest an intermediate architectural layer in which local noisy updates are organized into a distributed dynamical medium. Recent applications of quantum reservoir computing to realistic tasks and hardware platforms, including realized-volatility forecasting \cite{LiQRCFinance} and credit-default prediction on neutral-atom hardware \cite{VitaliQRCCredit}, reinforce the broader idea that structured noisy dynamics can be computationally useful. In the present setting, the combination of a persistent bosonic mode, tunable exchange dynamics, nonclassical state generation, Kerr-induced reshaping and engineered dissipation suggests a particularly rich hybrid reservoir architecture.

Beyond the many-body and open-system directions discussed above, the present architecture may also motivate future investigation in the context of quantum error correction. In particular, the genuinely hybrid use of the cavity mode may suggest routes toward bosonic-assisted syndrome extraction, hybrid qubit-boson encodings, or dissipative stabilization protocols \cite{Puri2019FaultTolerantSyndromeDetector,Gertler2021AutonomousQEC,Tsunoda2023BosonicEntanglingGates,Xu2024BosonicDVFaultTolerance}. Refining the experimental implementation of the proposed CNT cQED design and exploring quantitatively its use for QCA-inspired simulation, ancilla-based open-system dynamics, non-Gaussian many-body processing, reservoir-style quantum information processing and bosonic-assisted quantum error-correction strategies, are natural continuations of the present work.

\section{Acknowledgements}
This work is supported by the PEPR integrated project EPiQ ANR-22-PETQ-0007; by the ANR JCJC DisQC ANR- 22-CE47-0002-01 founded from the French National Re- search Agency; and the French government under the France 2030 investment plan, as part of the Initiative d’Excellence d’Aix-Marseille Université—A*MIDEX AMX-21-RID-011.
We thank Eoin Carolan for the reviewing of the manuscript.


\appendix


\section{Effective Hamiltonian and Gate Construction}
\label{app:supp_effham}

\subsection{Microscopic model and notation}

We consider two qubits longitudinally coupled to a microwave cavity mode, with a transverse inter-qubit exchange interaction and a weak Kerr nonlinearity,
\begin{equation}
\begin{split}
H_{S} &= H_{Q}+H_{c}\\
H_{Q} &:=
\sum_{i=1}^2 \frac{\omega_i}{2}\sigma_z^{(i)}
+ g(\sigma_z^{(1)}-\sigma_z^{(2)})(a+a^\dagger)
+ J\sigma_x^{(1)}\sigma_x^{(2)}, \\
H_{c}&:=\omega_c a^\dagger a
+ \chi (a^\dagger a)^2 .
\label{S:eq:H_full}
\end{split}
\end{equation}
We define the differential operator
\begin{equation}
Z:=\sigma_z^{(1)}-\sigma_z^{(2)},
\qquad
\Delta := \omega_1-\omega_2,
\qquad
\bar\omega:=\frac{\omega_1+\omega_2}{2},
\end{equation}
and decompose
\begin{equation}
\sum_{i=1}^2 \frac{\omega_i}{2}\sigma_z^{(i)}
=
\frac{\bar\omega}{2}\big(\sigma_z^{(1)}+\sigma_z^{(2)}\big)
+ \frac{\Delta}{4} Z.
\label{S:eq:qubit_split}
\end{equation}
The common-mode term $\propto (\sigma_z^{(1)}+\sigma_z^{(2)})$ is the zero operator in the single-excitation manifold
$\mathcal H_{\rm ap}=\mathrm{span}\{\ket{01},\ket{10}\}$ (``antiparallel'' manifold in the main text), and can be omitted
from the reduced problem without affecting the red-sideband derivation.

We also write
\begin{equation}
\label{eq:exch_term_supp}
\begin{split}
&\sigma_x^{(1)}\sigma_x^{(2)} = X + P, \\
&X:=\sigma_+^{(1)}\sigma_-^{(2)}+\sigma_-^{(1)}\sigma_+^{(2)}, \\
&P:=\sigma_+^{(1)}\sigma_+^{(2)}+\sigma_-^{(1)}\sigma_-^{(2)}.
\end{split}
\end{equation}
where we decomposed the exchange term $\sigma_{x}^{(1)}\sigma_{x}^{(2)}$ into the flip-flop term $X$ which only acts within the odd parity (parallel) sector $\mathrm{span}\{\ket{01},\ket{10}\}$, and $P$, only acting within the even parity (parallel) manifold $\mathrm{span}\{\ket{00},\ket{11}\}$. 

\subsection{Polaron transformation}
We apply the polaron unitary to the system Hamiltonian of Eq. ~\ref{S:eq:H_full}
\begin{equation}
U_P=e^S,\qquad S=\lambda Z(a^\dagger-a),\qquad \lambda=\frac{g}{\omega_c},
\label{S:eq:polaron_def}
\end{equation}
Using BCH,
\begin{equation}
U_P O U_P^\dagger = O+[S,O]+\frac{1}{2}[S,[S,O]]+\cdots.
\end{equation}
Choosing $\lambda=g/\omega_c$ cancels the linear cavity term in $H_C|_{\chi=0}+H_I$ and generates
\begin{equation}
H_{ZZ}= \frac{2g^2}{\omega_c}\,\sigma_z^{(1)}\sigma_z^{(2)}
\quad (\text{up to an irrelevant constant}).
\end{equation}
Using the decomposition of the exchange term of Eq. ~\ref{eq:exch_term_supp}, 
under $U_P$ the pair term $P$ is unchanged exactly, $U_PPU_P^\dagger=P$,
while the flip--flop term $X$ dresses as
\begin{equation}
U_P X U_P^\dagger
=
\sigma_+^{(1)}\sigma_-^{(2)} e^{+4\lambda(a^\dagger-a)}
+
\sigma_-^{(1)}\sigma_+^{(2)} e^{-4\lambda(a^\dagger-a)} .
\end{equation}
Expanding to first order in $\lambda$ yields
\begin{equation}
U_P X U_P^\dagger
\simeq
X + 4\lambda(a^\dagger-a)
\Big(\sigma_+^{(1)}\sigma_-^{(2)}-\sigma_-^{(1)}\sigma_+^{(2)}\Big),
\end{equation}
hence
\begin{equation}
H_{Q}^{P}:=U_{P}^{\dagger}SU_{P}\approx
\sum_{i=1}^2\frac{\omega_i}{2}\sigma_z^{(i)}
+\frac{2g^2}{\omega_c}\sigma_z^{(1)}\sigma_z^{(2)}
+J\sigma_x^{(1)}\sigma_x^{(2)}+H_{J}.
\label{S:eq:HSprime}
\end{equation}
where
\begin{equation}
H_{J}:=
4J\lambda(a^\dagger-a)
\Big(\sigma_+^{(1)}\sigma_-^{(2)}-\sigma_-^{(1)}\sigma_+^{(2)}\Big).
\label{S:eq:HJ1_def}
\end{equation}
In the weak-Kerr regime $\chi\ll\omega_c$ and low photon number, we neglect polaron-induced cavity and Kerr terms and adopt
$U_{P}^{\dagger}H_{c}U_{P}\approx H_{c}$.

%

\subsection{Diagonalization in the even and odd parity manifolds}
Since the exchange term $\sigma_{x}^{(1)}\sigma_{x}^{(2)}$ couples separately even and odd parity sectors and $H_{J}$ only has support on the odd parity sector, it is convenient to diagonalize the two sectors separately.

\subsubsection{Even (parallel) manifold}
Let $\Sigma:=\omega_1+\omega_2$. Up to a constant shift, the block is diagonalized by an angle $\theta_{\rm p}$,
\begin{equation}
\tan(2\theta_{\rm p})=\frac{2J}{\Sigma},\qquad
\Omega_{\rm p}=\sqrt{J^2+(\Sigma/2)^2},\qquad
\omega_{\rm p}=2\Omega_{\rm p},
\end{equation}
and we define dressed states
\begin{align}
\ket{P+}&=\cos\theta_{\rm p}\ket{00}+\sin\theta_{\rm p}\ket{11},\\
\ket{P-}&=-\sin\theta_{\rm p}\ket{00}+\cos\theta_{\rm p}\ket{11},
\end{align}
with energies $E_{P\pm}=\frac{2g^2}{\omega_c}\pm\Omega_{\rm p}$.
\subsubsection{Odd parity (antiparallel) manifold}
We introduce the Pauli operators
\begin{equation}
\begin{split}
&\tau_z := \ket{01}\!\bra{01}-\ket{10}\!\bra{10},
\qquad
\tau_x := \ket{10}\!\bra{01}+\ket{01}\!\bra{10}, \\
&\tau_y := i(\ket{01}\!\bra{10}-\ket{10}\!\bra{01}), \qquad
\tau_\pm := \frac{1}{2}(\tau_x \pm i\tau_y).
\end{split}
\end{equation}
Projecting the qubit Hamiltonian in the odd (antiparallel) sector, gives (up to an irrelevant constant shift)
\begin{equation}
H_{\rm ap} := \frac{\Delta}{2}\tau_z + J\tau_x+4J\lambda(a^\dagger-a)
i\tau_{y} \ .
\label{S:eq:H_ap_tau}
\end{equation}
We diagonalize Eq.~\eqref{S:eq:H_ap_tau} by a rotation about $\tau_y$,
\begin{equation}
\begin{split}
&R_{\rm ap}(\theta_{\rm ap}) = e^{-i\theta_{\rm ap}\tau_y},\\
&R_{\rm ap}^\dagger(\theta_{\rm ap})
\Big(\frac{\Delta}{2}\tau_z+J\tau_x\Big)
R_{\rm ap}(\theta_{\rm ap})
=
\Omega_{\rm ap}\,\tau_z,
\end{split}
\label{S:eq:Rap_def}
\end{equation}
where we define
\begin{equation}
\begin{split}
&\Omega_{\rm ap} := \sqrt{J^2+(\Delta/2)^2},
\qquad
\tan(2\theta_{\rm ap})=\frac{2J}{\Delta},\\
&\cos(2\theta_{\rm ap})=\frac{\Delta}{2\Omega_{\rm ap}}, \qquad \sin(2\theta_{\rm ap})=\frac{J}{\Omega_{\rm ap}}.
\end{split}
\label{S:eq:thetaap_defs}
\end{equation}

The dressed eigenstates in the antiparallel manifold are
\begin{equation}
\begin{split}
|AP+\rangle &= \cos\theta_{\mathrm{ap}} |01\rangle 
+ \sin\theta_{\mathrm{ap}} |10\rangle,
\\
|AP-\rangle &= -\sin\theta_{\mathrm{ap}} |01\rangle 
+ \cos\theta_{\mathrm{ap}} |10\rangle,
\end{split}
\end{equation}

with energies 
\begin{equation}
E_{\mathrm{AP}\pm}=-\frac{2g^2}{\omega_c}\pm\Omega_{\rm ap},
\quad
\omega_{\rm ap}=E_{\mathrm{AP}+}-E_{\mathrm{AP}-}=2\Omega_{\rm ap}.
\end{equation}
The dressed odd parity Hamiltonian writes
\begin{equation}
\begin{split}
&\mathcal{H}_{\rm ap}^{\prime}=
-\frac{2g^2}{\omega_c}\,\mathbb I_{\rm ap}
+\Omega_{\rm ap}\tilde{\tau}_z + \mathcal{H}_{\rm ap, J}^{\prime}, \\
&\mathcal{H}_{\rm ap, J}^{\prime}=
4iJ\lambda(a^\dagger-a)\tau_{y}.
\end{split}
\end{equation}
Here the dressed operators are defined as
\begin{equation}
\begin{split}
&\tilde{\tau}_z:=\ket{\mathrm{AP}+}\!\bra{\mathrm{AP}+}-\ket{\mathrm{AP}-}\!\bra{\mathrm{AP}-},
\qquad
\tilde{\tau}_+:=\ket{\mathrm{AP}+}\!\bra{\mathrm{AP}-},\\
&\tilde{\tau}_-:=\ket{\mathrm{AP}-}\!\bra{\mathrm{AP}+}.
\end{split}
\end{equation}
Since $\tau_y$ is invariant under the rotation $R_{\rm ap}$, one has
\begin{equation}
i\tau_y
=
iR_{\rm ap}^\dagger\tau_yR_{\rm ap}
=
\tilde{\tau}_- - \tilde{\tau}_+ .
\end{equation}
Accordingly,
\begin{equation}
\mathcal{H}_{\rm ap, J}^{\prime}
=
4J\lambda(a^\dagger-a)(\tilde{\tau}_- - \tilde{\tau}_+).
\end{equation}
In the following, we drop the tildes on the dressed operators to simplify the notation.

\subsection{Effective Hamiltonian in the dressed interaction picture}

We now move to the interaction picture with respect to
\begin{equation}
H_0:=\omega_c a^\dagger a+\mathcal{H}_{\rm ap}^{\prime},
\end{equation}
where the parallel sector Hamiltonian was omitted since it commutes through $H_J$ trivially. 
The cavity operators evolve as
\[
a_I(t)=ae^{-i\omega_c t},\qquad a_I^\dagger(t)=a^\dagger e^{i\omega_c t},
\]
and, using
\[
[\tau_z,\tau_\pm]=\pm 2\tau_\pm,
\]
the dressed ladder operators evolve as
\[
\tau_+(t)=\tau_+e^{i\omega_{\rm ap}t},
\qquad
\tau_-(t)=\tau_-e^{-i\omega_{\rm ap}t},
\]
hence
\begin{equation}
H_{J,I}(t)=
-4J\lambda
\big(a^\dagger e^{i\omega_c t}-a e^{-i\omega_c t}\big)
\big(\tau_+ e^{i\omega_{\rm ap}t}-\tau_- e^{-i\omega_{\rm ap}t}\big).
\end{equation}
Expanding,
\begin{align}
H_{J,I}(t)=-4J\lambda\Big[
& a^\dagger\tau_+\,e^{i(\omega_c+\omega_{\rm ap})t}
- a^\dagger\tau_-\,e^{i(\omega_c-\omega_{\rm ap})t}
\nonumber\\
&- a\tau_+\,e^{-i(\omega_c-\omega_{\rm ap})t}
+ a\tau_-\,e^{-i(\omega_c+\omega_{\rm ap})t}
\Big].
\end{align}

Near the resonance condition $\omega_c\simeq\omega_{\rm ap}$,
the terms oscillating at $\omega_c+\omega_{\rm ap}$ are rapidly rotating and are neglected in the rotating-wave approximation. Defining the detuning $\delta=\omega_c-\omega_{\rm ap}$,
we obtain the effective Hamiltonian
\begin{equation}
H_{BS}(t)
\simeq
+g_{\rm eff}
(
a^\dagger\tau_- e^{i\delta t}
+
a\tau_+ e^{-i\delta t})
\qquad g_{\rm eff}=4J\lambda=\frac{4Jg}{\omega_c}.
\end{equation}
%
\subsection{Bosonic beam-splitter gate and phase engineering}

We work with the effective Hamiltonian (interaction picture of $H_0$)
\begin{equation}
\begin{split}
&H(t)=H_{BS}(t)+H_K,
\qquad
H_{BS}= g_{\rm eff}\big(\tau_+ a + \tau_- a^\dagger\big),\\
&H_K=\chi (a^\dagger a)^2.
\end{split}
\label{S:eq:H_eff_rsb_kerr}
\end{equation}
The resulting gate is a composition of beam-splitter and Kerr evolutions, which in general do not factorize into two separate gate operations because $[H_{BS}(t),H_K]\neq 0$. Nevertheless, such a factorization can be recovered either perturbatively, when the Kerr phase accumulated during the beam-splitter step remains small over the occupied manifold, i.e.\ $|\chi|\,t_g\,n_{\max}\ll 1$, so that the dynamics may be treated within a first-order Zassenhaus expansion, or through an appropriate coherent control protocol. In the latter case, the resulting gate takes the form
\begin{equation}
\begin{split}
&U=U_{BS}\,U_K,
\qquad
U_{BS}:=\mathcal T\exp\!\Big(-i\!\int_0^{t_g}\!dt\,H_{BS}(t)\Big),\\
&U_K:=e^{-it_K \chi (a^\dagger a)^2},
\end{split}
\label{S:eq:U_target_def}
\end{equation}
In the following, for simplicity we just show the derivation of the $U_{BS}$ gate for the resonant case, i.e. $\delta=0$. 
\subsubsection{Resonant bosonic-swap gate ($\delta=0$)}
At resonance, $\delta=\omega_c-\omega_{\rm ap}=0$, the effective
qubit Hamiltonian reads
\begin{equation}
H_{BS}
=
g_{\rm eff}\left(a^\dagger\tau_-+a\tau_+\right),
\label{eq:H_rsb_supp}
\end{equation}
where
\begin{equation}
\tau_+=\ket{AP+}\!\bra{AP-},
\qquad
\tau_-=\ket{AP-}\!\bra{AP+}.
\end{equation}
The dressed even-parity states $\ket{P\pm}$ remain spectators under this
interaction.

The Hamiltonian \eqref{eq:H_rsb_supp} preserves invariant
two-dimensional subspaces of the form
\begin{equation}
\mathcal H_n=\mathrm{span}\Big\{\ket{AP+,n},\ket{AP-,n+1}\Big\},
\qquad n\ge 0,
\end{equation}
since
\begin{align}
H_{BS}\ket{AP+,n}
&=
g_{\rm eff}\sqrt{n+1}\,\ket{AP-,n+1},\\
H_{BS}\ket{AP-,n+1}
&=
g_{\rm eff}\sqrt{n+1}\,\ket{AP+,n}.
\end{align}
Thus, in the ordered basis
$\{\ket{AP+,n},\ket{AP-,n+1}\}$, one has
\begin{equation}
H_{BS}^{(n)}
=
g_{\rm eff}\sqrt{n+1}
\begin{pmatrix}
0 & 1\\
1 & 0
\end{pmatrix}
=
g_{\rm eff}\sqrt{n+1}\,\sigma_x .
\end{equation}
Exponentiating this block gives
\begin{equation}
U_n(t)=e^{-iH_{BS}^{(n)}t}
=
\cos\!\big(g_{\rm eff}t\sqrt{n+1}\big)\,\mathbb I
-
i\sin\!\big(g_{\rm eff}t\sqrt{n+1}\big)\,\sigma_x ,
\end{equation}
namely
\begin{equation}
U_n(t)=
\begin{pmatrix}
\cos\!\big(g_{\rm eff}t\sqrt{n+1}\big)
&
-i\sin\!\big(g_{\rm eff}t\sqrt{n+1}\big)
\\[2mm]
-i\sin\!\big(g_{\rm eff}t\sqrt{n+1}\big)
&
\cos\!\big(g_{\rm eff}t\sqrt{n+1}\big)
\end{pmatrix}.
\label{eq:Un_block_supp}
\end{equation}

Equivalently, the action of the propagator on each invariant block is
\begin{align}
U_{BS}(t)\ket{AP+,n}
&=
\cos\!\big(g_{\rm eff}t\sqrt{n+1}\big)\ket{AP+,n}
\nonumber\\
&\quad
-i\sin\!\big(g_{\rm eff}t\sqrt{n+1}\big)\ket{AP-,n+1},
\\
U_{BS}(t)\ket{AP-,n+1}
&=
\cos\!\big(g_{\rm eff}t\sqrt{n+1}\big)\ket{AP-,n+1}
\nonumber\\
&\quad
-i\sin\!\big(g_{\rm eff}t\sqrt{n+1}\big)\ket{AP+,n}.
\end{align}
Thus, the beam-splitter dynamics is block resolved,
with a photon-number-dependent Rabi angle originating from the bosonic
matrix elements
$a\ket{n}=\sqrt{n}\ket{n-1}$ and
$a^\dagger\ket{n}=\sqrt{n+1}\ket{n+1}$.

The same propagator can be written as a $4\times4$
block-operator matrix in the dressed basis
$\{\ket{P+},\ket{AP-},\ket{AP+},\ket{P-}\}$,
\begin{equation}
\begin{split}
U_{BS}(t)=
\begin{pmatrix}
\mathbb I & 0 & 0 & 0\\[2mm]
0 &
\cos\!\big(\Theta_{\hat N}\big) &
-i\,a^\dagger\,
\dfrac{\sin\!\big(\Theta_{\hat N+1}\big)}{\sqrt{\hat N+1}}
& 0\\[4mm]
0 &
-i\,a\,
\dfrac{\sin\!\big(\Theta_{\hat N}\big)}{\sqrt{\hat N}} &
\cos\!\big(\Theta_{\hat N+1}\big)
& 0\\[4mm]
0 & 0 & 0 & \mathbb I
\end{pmatrix},
\end{split}
\label{eq:U_BS_main}
\end{equation}
where $\hat N=a^\dagger a$ and
\begin{equation}
\Theta_{\hat N}=g_{\rm eff}t\sqrt{\hat N},
\qquad
\Theta_{\hat N+1}=g_{\rm eff}t\sqrt{\hat N+1}.
\end{equation}
This expression should be understood blockwise on the invariant
subspaces above; in particular, the apparent singularity of
$a/\sqrt{\hat N}$ at $\hat N=0$ is removed when the operator acts on the
relevant states $\ket{AP-,n+1}$.

The target coherent gate is then taken as the composition
\begin{equation}
U_{\rm target}=U_{BS}\,U_K,
\qquad
U_K=\exp\!\big[-it_K\chi(a^\dagger a)^2\big].
\end{equation}

\section{Open system dynamics}
\label{app:supp_lindblad}

\subsection{Total Hamiltonian, baths, and approximation scheme}

We start from
\begin{equation}
H_{\rm tot}=H_S+H_\gamma+H_{\gamma S}+H_\nu+H_{\nu S},
\end{equation}
with the system Hamiltonian split as
\begin{equation}
H_S=H_Q+H_C+H_I,
\end{equation}
where
\begin{align}
H_Q&=\sum_{i=1}^2\frac{\omega_i}{2}\sigma_z^{(i)}+J\,\sigma_x^{(1)}\sigma_x^{(2)},\\
H_C&=\omega_c a^\dagger a+\chi (a^\dagger a)^2,\\
H_I&=g\,Z\,(a+a^\dagger),\qquad Z:=\sigma_z^{(1)}-\sigma_z^{(2)}.
\end{align}

The photon bath, coupled to the cavity quadrature, is
\begin{align}
H_\gamma&=\sum_k\omega_k b_k^\dagger b_k,\\
H_{\gamma S}&=S_{\gamma}\otimes B_\gamma,\qquad
S_{\gamma}=a+a^\dagger,\\
B_\gamma&=\sum_k(f_k b_k^\dagger+f_k^\ast b_k),
\end{align}
while the qubit-sector bath, coupled directly to the qubits, is
\begin{align}
H_\nu&=\sum_q\omega_q c_q^\dagger c_q,\\
H_{\nu S}&=S_\nu\otimes B_\nu,\qquad
B_\nu=\sum_q(h_q c_q^\dagger+h_q^\ast c_q),\\
S_\nu&=\sum_{i=1}^2\big(\alpha_i\sigma_x^{(i)}+\beta_i\sigma_z^{(i)}\big).
\end{align}

We derive a Markovian master equation using the Davies weak-coupling limit, i.e.\ the Born--Markov and secular approximations \cite{BreuerPetruccione,Davies1974}. For each bath $x\in\{\gamma,\nu\}$ we introduce a spectral density $J_x(\omega)$ and thermal rates
\begin{equation}
\begin{split}    
&\gamma_x(\omega)=2\pi J_x(|\omega|)
\begin{cases}
n_B(|\omega|)+1,&\omega>0,\\
n_B(|\omega|),&\omega<0,
\end{cases}\\
&n_B(\omega)=\frac{1}{e^{\beta\omega}-1}.
\end{split}
\end{equation}
Given a system coupling operator $S$, the corresponding Davies eigenoperators are
\begin{equation}
S(\omega)=\sum_{\epsilon-\epsilon'=\omega}\Pi(\epsilon)\,S\,\Pi(\epsilon'),
\qquad
[H_0,S(\omega)]=-\omega\,S(\omega),
\end{equation}
where $H_0$ is the reference Hamiltonian used in the dissipative decomposition and $\Pi(\epsilon)$ projects onto its eigenspace of energy $\epsilon$. The Lindblad terms are then $\sum_\omega \gamma(\omega)\mathcal D[S(\omega)]$,
with dissipators $\mathcal{D}[L]\rho
=
L\rho L^\dagger
-
\frac12\{L^\dagger L,\rho\}$
and jump operators $L(\omega)=\sqrt{\gamma(\omega)}\,S(\omega)$,
where $\gamma(\omega)$ is the bath spectral rate evaluated at frequency $\omega$.

As in Sec.~\ref{app:supp_effham}, we simplify the system Hamiltonian by going to the polaron frame and diagonalizing separately the even and odd-parity manifolds. In the weak-Kerr regime $\chi\ll\omega_c$ and low photon number, we neglect polaron-induced cavity and Kerr terms and adopt
$U_{P}^{\dagger}H_{c}U_{P}\approx H_{c}$.

Throughout the dissipative construction we retain the exchange term in the coherent dynamics, but we do not when performing the eigenoperator decomposition. This is valid provided the splittings induced by the polaron-assisted hybridization remain unresolved by the baths.

\subsection{Born-Markov secular derivation}\label{sec:supp_lindblad_standard}
\subsubsection{Photon Lindblad operators}

In the polaron frame, the photon coupling operator reads
\begin{equation}
S_{\gamma}^{P}=U_P(a+a^\dagger)U_P^\dagger=(a+a^\dagger)-2\lambda Z.
\label{S:eq:polaron_quadrature}
\end{equation}

\paragraph{Cavity channels}

In the unresolved-Kerr regime, where the bath cannot resolve Kerr-split cavity transitions over the occupied manifold, the cavity dissipator is well approximated by the standard collapse operators
\begin{equation}
L_a=\sqrt{\kappa_\downarrow}\,a,
\qquad
L_{a^\dagger}=\sqrt{\kappa_\uparrow}\,a^\dagger,
\label{S:eq:cavity_jumps_supp}
\end{equation}
where
\begin{equation}
\kappa_\downarrow=\gamma_\gamma(\omega_c),
\qquad
\kappa_\uparrow=\gamma_\gamma(-\omega_c)
\end{equation}
are the photon emission and absorption rates.

If instead the Kerr splitting is resolved, the cavity part must be decomposed into the number-resolved transitions
\begin{equation}
a_n=\sqrt{n}\,\ket{n-1}\!\bra{n},
\qquad
\omega_n=\omega_c+\chi(2n-1),
\end{equation}
leading to the refined jump operators
\begin{equation}
L_{a_n}=\sqrt{\gamma_\gamma(\omega_n)}\,a_n,
\qquad
L_{a_n^\dagger}=\sqrt{\gamma_\gamma(-\omega_n)}\,a_n^\dagger.
\end{equation}
In the main text we work in the unresolved-Kerr regime.

\paragraph{Photon-induced dressed-qubit jumps}

Let $g_{\rm eff}:=4J\,g/\omega_c$ denote the strength of the polaron-induced exchange. In the derivation of the dissipator we assume that the additional hybridization induced by this term is unresolved by the photon bath, namely
\begin{equation}
g_{\rm eff}\sqrt{n_{\max}+1}\ll \kappa,
\label{S:eq:cond_B_clean}
\end{equation}
and more generally small compared with the spectral resolution of the bath. We therefore retain this term in the coherent dynamics, but not in the construction of the photon dissipator.

The operator $Z=\sigma_z^{(1)}-\sigma_z^{(2)}$ has support only on the antiparallel manifold. Using
\begin{equation}
\ket{\mathrm{AP}+}=c_{\rm ap}\ket{01}+s_{\rm ap}\ket{10},
\qquad
\ket{\mathrm{AP}-}=-s_{\rm ap}\ket{01}+c_{\rm ap}\ket{10},
\end{equation}
with $c_{\rm ap}=\cos\theta_{\rm ap}$ and $s_{\rm ap}=\sin\theta_{\rm ap}$, one obtains
\begin{align}
Z(+\omega_{\rm ap})
&=
-2\sin(2\theta_{\rm ap})\,
\ket{\mathrm{AP}-}\!\bra{\mathrm{AP}+},\\
Z(-\omega_{\rm ap})
&=
-2\sin(2\theta_{\rm ap})\,
\ket{\mathrm{AP}+}\!\bra{\mathrm{AP}-},\\
Z(0)
&=
2\cos(2\theta_{\rm ap})
\Big(\ket{\mathrm{AP}+}\!\bra{\mathrm{AP}+}
-\ket{\mathrm{AP}-}\!\bra{\mathrm{AP}-}\Big).
\label{S:eq:Z_decomp_AP_clean}
\end{align}

Since
\begin{equation}
S_\gamma^P=(a+a^\dagger)-2\lambda Z,
\end{equation}
the $Z$ component yields three photon-induced dressed-qubit channels in $\mathcal H_{\rm ap}$:
\begin{equation}
\begin{split}
L_{\gamma,{\rm ap}}(+\omega_{\rm ap})
&=
\sqrt{\gamma_\gamma(\omega_{\rm ap})}\,
\big(4\lambda\sin(2\theta_{\rm ap})\big)\,
\ket{\mathrm{AP}-}\!\bra{\mathrm{AP}+},
\\
L_{\gamma,{\rm ap}}(-\omega_{\rm ap})
&=
\sqrt{\gamma_\gamma(-\omega_{\rm ap})}\,
\big(4\lambda\sin(2\theta_{\rm ap})\big)\,
\ket{\mathrm{AP}+}\!\bra{\mathrm{AP}-},
\\
L_{\gamma,{\rm ap}}(0)
&=
\sqrt{\gamma_\gamma(0)}\,
\big(-4\lambda\cos(2\theta_{\rm ap})\big)\,\\
&\Big(\ket{\mathrm{AP}+}\!\bra{\mathrm{AP}+}
-\ket{\mathrm{AP}-}\!\bra{\mathrm{AP}-}\Big).
\end{split}
\label{S:eq:photon_induced_qubit_jumps_clean}
\end{equation}
\subsubsection{Phonon Lindblad operators}
In the polaron frame the qubit-sector bath coupling transforms as
\begin{equation}
S_\nu^P = U_P S_\nu U_P^\dagger,
\qquad
U_P=e^{\lambda Z(a^\dagger-a)},
\qquad
Z=\sigma_z^{(1)}-\sigma_z^{(2)}.
\end{equation}
Starting from
\begin{equation}
S_\nu=\sum_{i=1}^2\big(\alpha_i\sigma_x^{(i)}+\beta_i\sigma_z^{(i)}\big),
\label{S:eq:Snu_bare_clean}
\end{equation}
and expanding to first order in $\lambda=g/\omega_c$, one finds
\begin{equation}
S_\nu^P = S_\nu+[S,S_\nu]+\mathcal O(\lambda^2),
\qquad
S=\lambda Z(a^\dagger-a).
\end{equation}
Using
\begin{equation}
[Z,\sigma_z^{(i)}]=0,
\qquad
[Z,\sigma_x^{(1)}]=2i\sigma_y^{(1)},
\qquad
[Z,\sigma_x^{(2)}]=-2i\sigma_y^{(2)},
\end{equation}
this gives
\begin{equation}
[S,S_\nu]
=
2i\lambda(a^\dagger-a)
\big(\alpha_1\sigma_y^{(1)}-\alpha_2\sigma_y^{(2)}\big).
\end{equation}
Hence
\begin{equation}
S_\nu^P=S_\nu^{(0)}+S_\nu^{(1)}+\mathcal O(\lambda^2),
\label{S:eq:Snu_split_clean}
\end{equation}
with
\begin{align}
S_\nu^{(0)}
&=
\sum_{i=1}^2\big(\alpha_i\sigma_x^{(i)}+\beta_i\sigma_z^{(i)}\big),
\label{S:eq:Snu0_def_clean}
\\
S_\nu^{(1)}
&=
2i\lambda(a^\dagger-a)
\big(\alpha_1\sigma_y^{(1)}-\alpha_2\sigma_y^{(2)}\big).
\label{S:eq:Snu1_def_clean}
\end{align}

\paragraph{Zeroth-order qubit-sector bath dissipator}

At zeroth order in $\lambda$ we write
\begin{equation}
\begin{split}
&S_\nu^{(0)}=S_\nu^{(z)}+S_\nu^{(x)},
\qquad
S_\nu^{(z)}=\beta_1\sigma_z^{(1)}+\beta_2\sigma_z^{(2)}, \\
&S_\nu^{(x)}=\alpha_1\sigma_x^{(1)}+\alpha_2\sigma_x^{(2)}.
\end{split}
\end{equation}

\begin{itemize}
    \item \textbf{(i) $\beta$-terms: block-internal dressed relaxation and dephasing.}
    
    The operator $S_\nu^{(z)}$ acts separately within the parallel and antiparallel manifolds.

    In the parallel manifold,
    \begin{equation}
    S_{\nu,{\rm p}}^{(z)}
    =
    (\beta_1+\beta_2)
    \big(\ket{00}\!\bra{00}-\ket{11}\!\bra{11}\big),
    \end{equation}
    which yields
    \begin{equation}
    \begin{split}
    S_{\nu,{\rm p}}^{(z)}(+\omega_{\rm p})
    &=
    -(\beta_1+\beta_2)\sin(2\theta_{\rm p})\ket{P-}\!\bra{P+},
    \\
    S_{\nu,{\rm p}}^{(z)}(-\omega_{\rm p})
    &=
    -(\beta_1+\beta_2)\sin(2\theta_{\rm p})\ket{P+}\!\bra{P-},
    \\
    S_{\nu,{\rm p}}^{(z)}(0)
    &=
    (\beta_1+\beta_2)\cos(2\theta_{\rm p})\\
    &\Big(\ket{P+}\!\bra{P+}-\ket{P-}\!\bra{P-}\Big).
    \end{split}
    \end{equation}

    In the antiparallel manifold,
    \begin{equation}
    S_{\nu,{\rm ap}}^{(z)}
    =
    (\beta_1-\beta_2)
    \big(\ket{01}\!\bra{01}-\ket{10}\!\bra{10}\big),
    \end{equation}
    which yields
    \begin{equation}
    \begin{split}
    S_{\nu,{\rm ap}}^{(z)}(+\omega_{\rm ap})
    &=
    -(\beta_1-\beta_2)\sin(2\theta_{\rm ap})
    \ket{\mathrm{AP}-}\!\bra{\mathrm{AP}+},
    \\
    S_{\nu,{\rm ap}}^{(z)}(-\omega_{\rm ap})
    &=
    -(\beta_1-\beta_2)\sin(2\theta_{\rm ap})
    \ket{\mathrm{AP}+}\!\bra{\mathrm{AP}-},
    \\
    S_{\nu,{\rm ap}}^{(z)}(0)
    &=
    (\beta_1-\beta_2)\cos(2\theta_{\rm ap})\\
    &\Big(\ket{\mathrm{AP}+}\!\bra{\mathrm{AP}+}
    -\ket{\mathrm{AP}-}\!\bra{\mathrm{AP}-}\Big),
    \end{split}
    \end{equation}
    %

\item \textbf{(ii) $\alpha$-terms: cross-manifold transitions.}

    The operator
    \begin{equation}
    S_\nu^{(x)}=\alpha_1\sigma_x^{(1)}+\alpha_2\sigma_x^{(2)}
    \end{equation}
    flips one qubit and therefore connects the parallel and antiparallel manifolds. Defining the Bohr frequencies
    \begin{equation}
    \omega_{\mu\mu'}:=E_{{\rm AP}\mu}-E_{P\mu'},
    \qquad
    \mu,\mu'\in\{+,-\},
    \end{equation}
    and the overlaps
    \begin{equation}
    M^{(i)}_{\mu\mu'}=
    \braket{\mathrm{AP}\mu|\sigma_x^{(i)}|P\mu'},
    \qquad i\in\{1,2\},
    \end{equation}
    one finds
    \begin{equation}
    \begin{split}
    M^{(1)}_{++}&=c_{\rm ap}s_{\rm p}+s_{\rm ap}c_{\rm p}, \qquad
    M^{(2)}_{++}=c_{\rm ap}c_{\rm p}+s_{\rm ap}s_{\rm p},
    \\
    M^{(1)}_{-+}&=-s_{\rm ap}s_{\rm p}+c_{\rm ap}c_{\rm p},
    \quad
    M^{(2)}_{-+}=-s_{\rm ap}c_{\rm p}+c_{\rm ap}s_{\rm p},
    \\
    M^{(1)}_{+-}&=c_{\rm ap}c_{\rm p}-s_{\rm ap}s_{\rm p},
    \qquad
    M^{(2)}_{+-}=-c_{\rm ap}s_{\rm p}+s_{\rm ap}c_{\rm p},
    \\
    M^{(1)}_{--}&=-s_{\rm ap}c_{\rm p}-c_{\rm ap}s_{\rm p},
    \quad
    M^{(2)}_{--}=s_{\rm ap}s_{\rm p}+c_{\rm ap}c_{\rm p},
    \end{split}
    \end{equation}
    where $c_{\rm p}=\cos\theta_{\rm p}$, $s_{\rm p}=\sin\theta_{\rm p}$, $c_{\rm ap}=\cos\theta_{\rm ap}$, and $s_{\rm ap}=\sin\theta_{\rm ap}$.
    
    Defining
    \begin{equation}
    C_{\mu\mu'}=\alpha_1M^{(1)}_{\mu\mu'}+\alpha_2M^{(2)}_{\mu\mu'},
    \end{equation}
    the corresponding Davies jump operators are
    \begin{equation}
    L_{\nu,\mu\mu'}(\omega_{\mu\mu'})
    =
    \sqrt{\gamma_\nu(\omega_{\mu\mu'})}
    \,C_{\mu\mu'}\,
    \ket{\mathrm{AP}\mu}\!\bra{P\mu'},
    \label{S:eq:phonon_alpha_jumps_clean}
    \end{equation}
    with the reverse processes given by the Hermitian conjugates at $-\omega_{\mu\mu'}$.
\end{itemize}

\paragraph{First-order polaron correction to the qubit-sector bath coupling}

Retaining the first-order term $S_\nu^{(1)}$ introduces bath-assisted sideband transitions that simultaneously change the qubit manifold and the cavity occupation. Writing
\begin{equation}
Y_\alpha=\alpha_1\sigma_y^{(1)}-\alpha_2\sigma_y^{(2)},
\qquad
S_\nu^{(1)}=2i\lambda(a^\dagger-a)Y_\alpha,
\end{equation}
one finds transitions of the form
\begin{equation}
\ket{P\mu',n}\longleftrightarrow\ket{\mathrm{AP}\mu,n\pm1}.
\end{equation}
In the unresolved-Kerr regime the associated Bohr frequencies are
\begin{equation}
\omega_{\mu\mu'}\pm\omega_c,
\end{equation}
while in the resolved-Kerr regime they acquire the corresponding number-dependent shifts.

The eigenoperator components can be written as
\begin{align}
S_\nu^{(1)}(\omega_{\mu\mu'}+\omega_c)
&=
\sum_{n\ge0}
\sqrt{n+1}\,\widetilde{C}_{\mu\mu'}
\ket{\mathrm{AP}\mu,n+1}\!\bra{P\mu',n},
\\
S_\nu^{(1)}(\omega_{\mu\mu'}-\omega_c)
&=
-\sum_{n\ge1}
\sqrt{n}\,\widetilde{C}_{\mu\mu'}
\ket{\mathrm{AP}\mu,n-1}\!\bra{P\mu',n},
\end{align}
with reverse processes given by the Hermitian-conjugate operators at the opposite Bohr frequencies.

The coefficients are determined by the matrix elements
\begin{equation}
N^{(1)}_{\mu\mu'}=
\braket{\mathrm{AP}\mu|\sigma_y^{(1)}|P\mu'},
\qquad
N^{(2)}_{\mu\mu'}=
\braket{\mathrm{AP}\mu|\sigma_y^{(2)}|P\mu'},
\end{equation}
and
\begin{equation}
\widetilde C_{\mu\mu'}
=
2i\lambda
\Big(\alpha_1 N^{(1)}_{\mu\mu'}-\alpha_2 N^{(2)}_{\mu\mu'}\Big).
\end{equation}
Explicitly,
\begin{equation}
\begin{split}
\widetilde C_{++}
&=
-2\lambda\Big[
\alpha_1(s_{\rm ap}c_{\rm p}-c_{\rm ap}s_{\rm p})
-
\alpha_2(c_{\rm ap}c_{\rm p}-s_{\rm ap}s_{\rm p})
\Big],
\\
\widetilde C_{-+}
&=
-2\lambda\Big[
\alpha_1(s_{\rm ap}s_{\rm p}+c_{\rm ap}c_{\rm p})
+
\alpha_2(s_{\rm ap}c_{\rm p}+c_{\rm ap}s_{\rm p})
\Big],
\\
\widetilde C_{+-}
&=
2\lambda\Big[
\alpha_1(c_{\rm ap}c_{\rm p}+s_{\rm ap}s_{\rm p})
-
\alpha_2(c_{\rm ap}s_{\rm p}+s_{\rm ap}c_{\rm p})
\Big],
\\
\widetilde C_{--}
&=
-2\lambda\Big[
\alpha_1(s_{\rm ap}c_{\rm p}-c_{\rm ap}s_{\rm p})
-
\alpha_2(s_{\rm ap}s_{\rm p}-c_{\rm ap}c_{\rm p})
\Big].
\end{split}
\end{equation}

The corresponding Lindblad operators are then obtained by multiplying these eigenoperators by the appropriate square-root bath rates. For example,
\begin{equation}
L_{\nu,\mu\mu'}^{(+)}
=
\sqrt{\gamma_\nu(\omega_{\mu\mu'}+\omega_c)}
\sum_{n\ge0}
\sqrt{n+1}\,\widetilde C_{\mu\mu'}\,
\ket{\mathrm{AP}\mu,n+1}\!\bra{P\mu',n},
\end{equation}
with analogous expressions for the channels at $\omega_{\mu\mu'}-\omega_c$ and for the reverse processes.
\subsection{Collision-model derivation}
\label{app:supp_collision}

In this subsection we show that the same effective Lindblad dynamics derived above from the Davies weak-coupling limit also arises from a collision-model construction \cite{CiccarelloCollisionReview,CusumanoCollisionGuide}.

\subsubsection{Weak-collision scaling}

We model the environment as a sequence of independent ancillas, refreshed at each time step $\Delta t$. During one collision step, the dressed qubit-cavity system interacts with fresh photon and qubit-sector ancillas according to the joint unitary

\begin{equation}
U_{\Delta t}
=
\exp\!\left[
-i\Delta t\,H_{\rm eff}
-i\sqrt{\Delta t}\sum_{r=\gamma,\nu}V_r
\right],
\label{S:eq:coll_unitary_clean}
\end{equation}
where $H_{\rm eff}$ is the same coherent effective Hamiltonian derived in section ~\ref{app:supp_effham} and used in the standard open-system treatment of section ~\ref{sec:supp_lindblad_standard}, while $V_\gamma$ and $V_\nu$ are the collision couplings associated with the photon and qubit-sector ancilla streams (see below).
As in the Davies derivation, the collision operators are built from the same dressed eigenoperators $S_r(\omega)$ obtained from the same reference Hamiltonian $H_0$ entering the dissipative decomposition. In particular, the cavity-assisted exchange term $H_J$ is retained in the coherent dynamics but neglected in the eigenoperator construction, and the cavity bath is treated within the same unresolved-Kerr approximation. The $\sqrt{\Delta t}$ scaling is the standard weak-collision scaling ensuring that the coherent contribution remains of order $\Delta t$ while the dissipative contribution survives with a finite limit as $\Delta t\to0$ \cite{AttalPautrat,CiccarelloCollisionReview}.

The reduced system state after one collision is therefore
\begin{equation}
\Phi_{\Delta t}(\rho)
=
\Tr_{\rm anc}
\!\left[
U_{\Delta t}
(\rho\otimes \eta)
U_{\Delta t}^\dagger
\right],
\label{S:eq:coll_map_clean}
\end{equation}
with $\eta=\eta_\gamma\otimes\eta_\nu$ the state of the incoming ancillas. We assume that the ancillas are identically prepared and independent from one step to the next.

For each bath label $r\in\{\gamma,\nu\}$, we choose
\begin{equation}
V_r=
\sum_{\omega>0}
\Big(
S_r(\omega)\otimes b_{r,\omega}^\dagger
+
S_r^\dagger(\omega)\otimes b_{r,\omega}
\Big)
+
S_r(0)\otimes X_r,
\label{S:eq:Vr_collision_clean}
\end{equation}
where $b_{r,\omega}$ and $X_r=X_r^\dagger$ act on the fresh ancilla of stream $r$. We further assume vanishing ancilla first moments,
\begin{equation}
\Tr(\eta_r\,b_{r,\omega})=0,
\qquad
\Tr(\eta_r\,X_r)=0,
\label{S:eq:anc_first_moments}
\end{equation}
so that only second-order terms contribute to the dissipative continuum limit.

\subsubsection{Continuum limit}

Expanding Eq.~\eqref{S:eq:coll_unitary_clean} up to order $\Delta t$ gives
\begin{equation}
U_{\Delta t}
=
\mathbb I
-i\Delta t\,H_{\rm eff}
-i\sqrt{\Delta t}\sum_r V_r
-\frac{\Delta t}{2}\sum_{r,r'}V_rV_{r'}
+
O(\Delta t^{3/2}).
\end{equation}
Substituting into Eq.~\eqref{S:eq:coll_map_clean}, the terms of order $\sqrt{\Delta t}$ vanish after tracing over the ancillas because of Eq.~\eqref{S:eq:anc_first_moments}. The surviving terms of order $\Delta t$ then yield
\begin{equation}
\Phi_{\Delta t}(\rho)-\rho
=
-i\Delta t[H_{\rm eff}+H_{\rm LS},\rho]
+
\Delta t\,\mathcal L_{\rm coll}(\rho)
+
O(\Delta t^{3/2}),
\label{S:eq:discrete_to_cont_clean}
\end{equation}
where $H_{\rm LS}$ is a Lamb-shift term generated by ancilla correlations.

For the collision Hamiltonian \eqref{S:eq:Vr_collision_clean}, the dissipative contribution takes the standard GKLS form \cite{AttalPautrat,CiccarelloCollisionReview}
\begin{equation}
\begin{split}
\mathcal L_{\rm coll}(\rho)
&=
\sum_{r,\omega>0}
\Gamma_r^\downarrow(\omega)\,
\mathcal D[S_r(\omega)]\rho
+
\sum_{r,\omega>0}
\Gamma_r^\uparrow(\omega)\,
\mathcal D[S_r^\dagger(\omega)]\rho\\
&+
\sum_r
\Gamma_r^0\,
\mathcal D[S_r(0)]\rho,
\label{S:eq:Lcoll_general_clean}
\end{split}
\end{equation}
with effective rates fixed by ancilla second moments,
\begin{align}
\Gamma_r^\downarrow(\omega)
&=
\Tr\!\big(\eta_r\,b_{r,\omega}b_{r,\omega}^\dagger\big),\\
\Gamma_r^\uparrow(\omega)
&=
\Tr\!\big(\eta_r\,b_{r,\omega}^\dagger b_{r,\omega}\big),\\
\Gamma_r^0
&=
\Tr\!\big(\eta_r\,X_r^2\big),
\end{align}
up to the normalization convention adopted in Eq.~\eqref{S:eq:Vr_collision_clean}. Passing to the continuous time limit $t=n\Delta t$ gives
\begin{equation}
\begin{split}
&\dot\rho
=
-i[H_{\rm eff}+H_{\rm LS},\rho]
+
\sum_{r,\omega>0}
\Gamma_r^\downarrow(\omega)\,
\mathcal D[S_r(\omega)]\rho
\\
&+\sum_{r,\omega>0}
\Gamma_r^\uparrow(\omega)\,
\mathcal D[S_r^\dagger(\omega)]\rho
+
\sum_r
\Gamma_r^0\,
\mathcal D[S_r(0)]\rho.
\end{split}
\label{S:eq:GKLS_collision_clean}
\end{equation}

The continuum-limit collision-model master equation \eqref{S:eq:GKLS_collision_clean} has the same GKLS structure as the Davies master equation derived above. The equivalence becomes explicit upon identifying the ancilla correlators with the corresponding bath spectral rates,
\begin{equation}
\Gamma_r^\downarrow(\omega)=\gamma_r(\omega),
\qquad
\Gamma_r^\uparrow(\omega)=\gamma_r(-\omega),
\qquad
\Gamma_r^0=\gamma_r(0).
\label{S:eq:Gamma_equals_gamma_clean}
\end{equation}

\section{Average Gate Fidelity in the Weak-Dissipation Regime}
\label{app:supp_fidelity}

We compute the average gate fidelity of the implemented evolution over a gate time $t_g$
in the weak-dissipation regime $\gamma_k t_g\ll 1$, using the Lindblad operators $L_k$
derived in the previous section (polaron frame, weak-Kerr approximation), without explicitly
constructing the full superoperator.

\subsection{Truncated Hilbert space and projected operators}
All Haar averages and traces are evaluated on the same finite-dimensional Hilbert space used
in the numerics,
\begin{equation}
\begin{split}
&\mathcal H_d=\mathcal H_{\rm qb}\otimes \mathcal H_{\rm cav}^{(n_{\max})},
\qquad
\mathcal H_{\rm cav}^{(n_{\max})}=\mathrm{span}\{\ket{n}\}_{n=0}^{n_{\max}},\\
&d=\dim\mathcal H_d=4(n_{\max}+1).
\end{split}
\label{S:eq:Hd_def}
\end{equation}

\paragraph{Remark (truncation and ``unitarity'').}
A simplification below uses invariance of the trace under unitary conjugation, which is exact
provided the ideal propagator $U(t)$ is unitary on $\mathcal H_d$ (i.e.\ negligible leakage beyond $n_{\max}$).
If leakage is non-negligible, one should use the time-dependent expression
Eq.~\eqref{S:eq:Favg_general_time} (evaluated along the ideal trajectory), or increase $n_{\max}$ until convergence.

For cavity operators we use the projected convention
\begin{equation}
P:=\sum_{n=0}^{n_{\max}}\ket{n}\!\bra{n},
\qquad
a_P:=PaP,\qquad a_P^\dagger:=Pa^\dagger P,
\end{equation}
which ensures complete positivity on $\mathcal H_d$. Identities such as
$a a^\dagger=a^\dagger a+\mathbb I$ are only exact before truncation; boundary corrections
vanish when the cutoff state is unoccupied.

\subsection{General weak-dissipation formula for $F_{\rm avg}$}
We start from the Lindblad equation
\begin{equation}
\begin{split}
&\dot\rho(t)=\mathcal L_0[\rho(t)]+\mathcal L_D[\rho(t)],
\quad
\mathcal L_0[\rho]=-i[H(t),\rho],\\
&\mathcal L_D[\rho]=\sum_k \mathcal D[L_k]\rho,
\end{split}
\end{equation}
with $\mathcal D[L]\rho=L\rho L^\dagger-\tfrac12\{L^\dagger L,\rho\}$.

To first order in the rates, we evaluate $\mathcal L_D$ along the ideal unitary trajectory
$\rho_0(t)=U(t)\rho(0)U^\dagger(t)$, giving (in the ideal frame)
\begin{equation}
\begin{split}
&U_g^\dagger \mathcal E(\rho)\,U_g
\simeq
\rho + \sum_k \int_0^{t_g} dt\; \mathcal D[\tilde L_k(t)]\rho,\\
&\tilde L_k(t):=U^\dagger(t)L_kU(t),\\
&U_g:=U(t_g).
\end{split}
\end{equation}
The average gate fidelity is \cite{Nielsen}
\begin{equation}
F_{\rm avg}=\int d\psi\;\bra{\psi}\,U_g^\dagger\mathcal E(\ket{\psi}\!\bra{\psi})U_g\,\ket{\psi}.
\end{equation}
Using standard Haar identities yields the first-order expression
\begin{equation}
\begin{split}
F_{\rm avg}&=
1-\frac{1}{d+1}\sum_k\int_0^{t_g}\!dt\,
\left[
\Tr\!\big(\tilde L_k^\dagger(t)\tilde L_k(t)\big)\right.\\
&\left.-\frac{1}{d}\big|\Tr(\tilde L_k(t))\big|^2
\right]+\mathcal O(\gamma^2 t_g^2).
\end{split}
\label{S:eq:Favg_general_time}
\end{equation}
If $U(t)$ is unitary on $\mathcal H_d$, then by cyclicity of the trace
\begin{equation}
\Tr\!\big(\tilde L_k^\dagger(t)\tilde L_k(t)\big)=\Tr(L_k^\dagger L_k),
\qquad
\Tr(\tilde L_k(t))=\Tr(L_k),
\end{equation}
and Eq.~\eqref{S:eq:Favg_general_time} simplifies to
\begin{equation}
F_{\rm avg}
=
1-\frac{t_g}{d+1}\sum_k
\left[
\Tr(L_k^\dagger L_k)-\frac{1}{d}|\Tr(L_k)|^2
\right]
+\mathcal O(\gamma^2 t_g^2).
\label{S:eq:Favg_simple}
\end{equation}

In our case, all jump operators are traceless on $\mathcal H_d$ under the projected-operator convention:
$\Tr(a_P)=\Tr(a_P^\dagger)=0$ and the qubit dressed-basis operators are either off-diagonal projectors
or differences of projectors. Hence $\Tr(L_k)=0$ and
\begin{equation}
F_{\rm avg}
=
1-\frac{t_g}{d+1}\sum_k \Tr(L_k^\dagger L_k)
+\mathcal O(\gamma^2 t_g^2).
\label{S:eq:Favg_working_traceless}
\end{equation}

Moreover, we note that, if $L=L_q\otimes \mathbb I_c$ is qubit-only, then
\begin{equation}
\Tr(L^\dagger L)=\Tr_q(L_q^\dagger L_q)\,\Tr_c(\mathbb I_c)=(n_{\max}+1)\Tr_q(L_q^\dagger L_q).
\label{S:eq:trace_factorization}
\end{equation}

Finally, we separate photon and qubit-sector bath contributions in Eq. (\ref{S:eq:Favg_working_traceless}) and write
\begin{equation}
F_{\rm avg}=1-(F_\gamma+F_\nu)+\mathcal O(\gamma^2 t_g^2).
\label{S:eq:def_Fgamma_Fnu}
\end{equation}
where 
\begin{equation}
F_\gamma
:=
\frac{t_g}{d+1}\sum_{k\in\gamma}\Tr(L_k^\dagger L_k) \qquad 
F_\nu
:=
\frac{t_g}{d+1}\sum_{k\in\nu}\Tr(L_k^\dagger L_k).
\end{equation}
\subsection{Photon contribution}
\paragraph{(i) Cavity loss/absorption.}

Using $L_a=\sqrt{\gamma_\gamma(\omega_c)}\,a_P$ and $L_{a^\dagger}=\sqrt{\gamma_\gamma(-\omega_c)}\,a_P^\dagger$,
\begin{equation}
\begin{split}
\Tr(L_a^\dagger L_a)
&=
\gamma_\gamma(\omega_c)\,\Tr_q(\mathbb I_q)\Tr_c(a_P^\dagger a_P)
\\
&=\gamma_\gamma(\omega_c)\cdot 4\sum_{n=0}^{n_{\max}}n=2\,\gamma_\gamma(\omega_c)\,n_{\max}(n_{\max}+1),
\\
\Tr(L_{a^\dagger}^\dagger L_{a^\dagger})
&=
\gamma_\gamma(-\omega_c)\cdot 4\sum_{n=0}^{n_{\max}-1}(n+1)\\
&=2\,\gamma_\gamma(-\omega_c)\,n_{\max}(n_{\max}+1).
\end{split}
\end{equation}

\paragraph{(ii) Photon-induced antiparallel dressed channels.}
Photon-induced qubit dissipation comes from the $-2\lambda Z$ part of $S_\gamma=(a+a^\dagger)-2\lambda Z$ and
acts only in the antiparallel manifold spanned by $\{\ket{\mathrm{AP}\pm}\}$. 
The jump operators from the Lindblad derivation are given by Eq. ~\ref{S:eq:photon_induced_qubit_jumps_clean}
and since they are qubit-only we use Eq.~\eqref{S:eq:trace_factorization} to get
\begin{equation}
\begin{split}
&\sum_{\ell\in\{+\omega_{\rm ap},-\omega_{\rm ap},0\}}
\Tr\!\left(L_{\gamma,{\rm ap}}(\ell)^\dagger L_{\gamma,{\rm ap}}(\ell)\right)\\
&=
(n_{\max}+1)\Big[
16\lambda^2\sin^2(2\theta_{\rm ap})\big(\gamma_\gamma(\omega_{\rm ap})+\gamma_\gamma(-\omega_{\rm ap})\big)\\
&+32\lambda^2\cos^2(2\theta_{\rm ap})\gamma_\gamma(0)
\Big].
\end{split}
\label{S:eq:Tr_photon_ap}
\end{equation}
where for instance
\begin{equation}
\begin{split}    
&\Tr(L_{\gamma,{\rm ap}}(+\omega_{\rm ap})^\dagger L_{\gamma,{\rm ap}}(+\omega_{\rm ap}))
\\
&=\Tr_q\!\left(\gamma_\gamma(\omega_{\rm ap})(4\lambda\sin 2\theta_{\rm ap})^2\,\ket{\mathrm{AP}+}\!\bra{\mathrm{AP}+}\right)\Tr_c(\mathbb I_c)
\nonumber\\
&=
\gamma_\gamma(\omega_{\rm ap})(4\lambda\sin 2\theta_{\rm ap})^2\cdot 1 \cdot (n_{\max}+1)
\nonumber\\
&=(n_{\max}+1)\gamma_\gamma(\omega_{\rm ap})\,16\lambda^{2}\sin^{2}(2\theta_{\rm ap}) .
\end{split}
\end{equation}

\paragraph{Photon trace sum.}
Collecting cavity and photon-induced qubit terms,
\begin{equation}
\begin{split} 
&\sum_{k\in\gamma}\Tr(L_k^\dagger L_k)
=
\;2n_{\max}(n_{\max}+1)\big(\gamma_\gamma(\omega_c)+\gamma_\gamma(-\omega_c)\big)
\\
&+(n_{\max}+1)\Big[
16\lambda^2\sin^2(2\theta_{\rm ap})\big(\gamma_\gamma(\omega_{\rm ap})+\gamma_\gamma(-\omega_{\rm ap})\big)\\
&+32\lambda^2\cos^2(2\theta_{\rm ap})\gamma_\gamma(0)
\Big].
\end{split}
\label{S:eq:Tr_photon_total}
\end{equation}

\subsection{Qubit-sector bath contribution $F_\nu$ (block-internal + cross-manifold channels)}
The qubit-sector bath is coupled through
\begin{equation}
H_{\nu S}=S_\nu\otimes B_\nu,
\qquad
S_\nu=\sum_{i=1}^2\big(\alpha_i\sigma_x^{(i)}+\beta_i\sigma_z^{(i)}\big) \ .
\label{S:eq:Snudef}
\end{equation} 

\paragraph{(i) Block-internal $\beta$-channels.}
Using the dressed-basis qubit-sector bath Lindbladians from the previous section, one finds
\begin{equation}
\begin{aligned}
&\sum_{\ell\in\{\pm\omega_{\rm p},0\}}
\Tr\!\left(L_{\nu,{\rm p}}(\ell)^\dagger L_{\nu,{\rm p}}(\ell)\right)=(n_{\max}+1)(\beta_1+\beta_2)^2\\
&\Big[
\sin^2(2\theta_{\rm p})\big(\gamma_\nu(\omega_{\rm p})+\gamma_\nu(-\omega_{\rm p})\big)
+2\cos^2(2\theta_{\rm p})\gamma_\nu(0)
\Big],
\\
&\sum_{\ell\in\{\pm\omega_{\rm ap},0\}}
\Tr\!\left(L_{\nu,{\rm ap}}(\ell)^\dagger L_{\nu,{\rm ap}}(\ell)\right)=
(n_{\max}+1)(\beta_1-\beta_2)^2\\
&\Big[
\sin^2(2\theta_{\rm ap})\big(\gamma_\nu(\omega_{\rm ap})+\gamma_\nu(-\omega_{\rm ap})\big)
+2\cos^2(2\theta_{\rm ap})\gamma_\nu(0)
\Big].
\end{aligned}
\label{S:eq:Tr_phonon_beta_total}
\end{equation}

\paragraph{(ii) Cross-manifold $\alpha$-channels.}
Writing the cross-manifold channels compactly as
\begin{equation}
\begin{split}
L_{\nu,\mu\mu^{\prime}}(\omega_{\mu\mu^{\prime}})
=
&\sqrt{\gamma_\nu(\omega_{\mu\mu^{\prime}})}
\Big(\alpha_1 M^{(1)}_{\mu\mu^{\prime}}+\alpha_2 M^{(2)}_{\mu\mu^{\prime}}\Big)\,\\
&\ket{\mathrm{AP}\mu}\!\bra{P\mu^{\prime}},
\qquad \mu,\mu^{\prime}\in\{\pm\},
\end{split}
\end{equation}
their contribution to $\sum_{k\in\nu}\Tr(L_k^\dagger L_k)$ is
\begin{equation}
(n_{\max}+1)\sum_{\mu,\mu^{\prime}=\pm}
\Big|\alpha_1 M^{(1)}_{\mu\mu^{\prime}}+\alpha_2 M^{(2)}_{\mu\mu^{\prime}}\Big|^2
\big(\gamma_\nu(\omega_{\mu\mu^{\prime}})+\gamma_\nu(-\omega_{\mu\mu^{\prime}})\big).
\label{S:eq:Tr_phonon_cross_total}
\end{equation}
For instance, we compute the explicit contribution of
\begin{equation}
L \equiv L_{\nu,++}(+\omega_{++})
=
\sqrt{\gamma_\nu(\omega_{++})}\,C_{++}\,\ket{\mathrm{AP}+}\!\bra{\mathrm{P}+}.
\end{equation}
where
\begin{equation}
C_{++}
\equiv
\alpha_1(c_{\rm ap}s_{\rm p}+s_{\rm ap}c_{\rm p})
+\alpha_2(c_{\rm ap}c_{\rm p}+s_{\rm ap}s_{\rm p}),
\end{equation}
with $c_{\rm p}=\cos\theta_{\rm p}$, $s_{\rm p}=\sin\theta_{\rm p}$,
$c_{\rm ap}=\cos\theta_{\rm ap}$, and $s_{\rm ap}=\sin\theta_{\rm ap}$.
Then
\begin{equation}
L^\dagger L
=
\gamma_\nu(\omega_{++})|C_{++}|^2\,
\ket{\mathrm{P}+}\!\bra{\mathrm{P}+}.
\end{equation}
This is qubit-only $\otimes\,\mathbb I_c$, hence
\begin{align}
\Tr(L^\dagger L)
&=
\Tr_q\!\left(\gamma_\nu(\omega_{++})|C_{++}|^2\ket{\mathrm{P}+}\!\bra{\mathrm{P}+}\right)
\Tr_c(\mathbb I_c)
\nonumber\\
&=(n_{\max}+1)\,\gamma_\nu(\omega_{++})\,|C_{++}|^2.
\label{eq:Tr_cross_first}
\end{align}

Similarly for $L_-=\sqrt{\gamma_\nu(-\omega_{++})}\,C_{++}^\ast\ket{\mathrm{P}+}\!\bra{\mathrm{AP}+}$,
one finds
\begin{equation}
\Tr(L_-^\dagger L_-)
=
(n_{\max}+1)\,\gamma_\nu(-\omega_{++})\,|C_{++}|^2.
\end{equation}
Thus, since each cross-manifold jump $L_{\nu,\mu\nu}(\pm\omega_{\mu\nu})$ is qubit-only 
\begin{equation}
\begin{split}
&\Tr\!\left(L_{\nu,\mu\mu^{\prime}}(\pm\omega_{\mu\mu^{\prime}})^\dagger L_{\nu,\mu\mu^{\prime}}(\pm\omega_{\mu\mu^{\prime}})\right)=\\
&(n_{\max}+1)\,\gamma_\nu(\pm\omega_{\mu\mu^{\prime}})\,|C_{\mu\mu^{\prime}}|^2.
\end{split}
\end{equation}
Summing absorption and emission for fixed $(\mu,\mu^{\prime})$ gives
\begin{equation}
\begin{split}
\Sigma_{\nu,\mu\mu^{\prime}}^{(\alpha)}
&\equiv
\Tr\!\left(L_{\nu,\mu\mu^{\prime}}(+\omega_{\mu\mu^{\prime}})^\dagger L_{\nu,\mu\mu^{\prime}}(+\omega_{\mu\mu^{\prime}})\right)\\
&+
\Tr\!\left(L_{\nu,\mu\mu^{\prime}}(-\omega_{\mu\mu^{\prime}})^\dagger L_{\nu,\mu\mu^{\prime}}(-\omega_{\mu\mu^{\prime}})\right)\\
&=(n_{\max}+1)\,|C_{\mu\mu^{\prime}}|^2\Big[\gamma_\nu(\omega_{\mu\mu^{\prime}})+\gamma_\nu(-\omega_{\mu\mu^{\prime}})\Big].
\end{split}
\end{equation}

Thus, the total fidelity contribution coming from all cross-manifold qubit-sector bath channels is
\begin{equation}
\begin{split}
F_{\nu}^{(\alpha)}
&=
\frac{t_g}{d+1}
\sum_{\mu,\mu^{\prime}=\pm}
\Sigma_{\nu,\mu\mu^{\prime}}^{(\alpha)}=
\frac{t_g}{4(n_{\max}+1)+1}\, \\
&(n_{\max}+1)\sum_{\mu,\mu^{\prime}=\pm}
|C_{\mu\mu^{\prime
}}|^2\Big[\gamma_\nu(\omega_{\mu\mu^{\prime}})+\gamma_\nu(-\omega_{\mu\mu^{\prime}})\Big].
\end{split}
\label{eq:F_alpha_total}
\end{equation}
\paragraph{(iii) First-order polaron correction to the qubit-sector bath contribution.}
If the qubit-sector bath coupling operator is retained to first order in the
polaron parameter, one must also include
\begin{equation}
S_\nu^{(1)}
=
2i\lambda(a^\dagger-a)
\big(\alpha_1\sigma_y^{(1)}-\alpha_2\sigma_y^{(2)}\big).
\end{equation}
As shown in the Lindblad derivation, this term generates cavity-assisted
cross-manifold transitions
\begin{equation}
\ket{P\mu',n}\longleftrightarrow\ket{\mathrm{AP}\mu,n\pm1},
\qquad
\mu,\mu'\in\{+,-\},
\end{equation}
with Bohr frequencies
\begin{equation}
\omega_{\mu\mu'}\pm\omega_c,
\qquad
\omega_{\mu\mu'}:=E_{\mathrm{AP}\mu}-E_{P\mu'} .
\end{equation}

The corresponding eigenoperator components are
\begin{align}
S_\nu^{(1)}(\omega_{\mu\mu'}+\omega_c)
&=
\sum_{n=0}^{n_{\max}-1}
\sqrt{n+1}\,\widetilde C_{\mu\mu'}
\ket{\mathrm{AP}\mu,n+1}\!\bra{P\mu',n},
\\
S_\nu^{(1)}(\omega_{\mu\mu'}-\omega_c)
&=
-\sum_{n=1}^{n_{\max}}
\sqrt{n}\,\widetilde C_{\mu\mu'}
\ket{\mathrm{AP}\mu,n-1}\!\bra{P\mu',n},
\end{align}
where the coefficients $\widetilde C_{\mu\mu'}$ are given in the
previous subsection.

Hence the associated Lindblad operators read
\begin{align}
L_{\nu,\mu\mu'}^{(+)}
&=
\sqrt{\gamma_\nu(\omega_{\mu\mu'}+\omega_c)}
\sum_{n=0}^{n_{\max}-1}
\sqrt{n+1}\,\widetilde C_{\mu\mu'}
\ket{\mathrm{AP}\mu,n+1}\!\bra{P\mu',n},
\\
L_{\nu,\mu\mu'}^{(-)}
&=
\sqrt{\gamma_\nu(\omega_{\mu\mu'}-\omega_c)}
\sum_{n=1}^{n_{\max}}
\big(-\sqrt{n}\,\widetilde C_{\mu\mu'}\big)
\ket{\mathrm{AP}\mu,n-1}\!\bra{P\mu',n}.
\end{align}

For example, for
\begin{equation}
\begin{split}
L\equiv L_{\nu,++}^{(+)}
&=
\sqrt{\gamma_\nu(\omega_{++}+\omega_c)}
\sum_{n=0}^{n_{\max}-1}
\sqrt{n+1}\,\widetilde C_{++}\, \\
&\ket{\mathrm{AP}+,n+1}\!\bra{P+,n},
\end{split}
\end{equation}
one finds
\begin{align}
L^\dagger L
&=
\gamma_\nu(\omega_{++}+\omega_c)\,|\widetilde C_{++}|^2
\sum_{n=0}^{n_{\max}-1}(n+1)\,
\ket{P+,n}\!\bra{P+,n}.
\end{align}
Therefore
\begin{align}
\Tr(L^\dagger L)
&=
\gamma_\nu(\omega_{++}+\omega_c)\,|\widetilde C_{++}|^2
\sum_{n=0}^{n_{\max}-1}(n+1)
\nonumber\\
&=
\gamma_\nu(\omega_{++}+\omega_c)\,|\widetilde C_{++}|^2
\frac{n_{\max}(n_{\max}+1)}{2}.
\label{eq:Tr_first_order_example}
\end{align}

Similarly, for the channel at frequency $\omega_{\mu\mu'}+\omega_c$ one obtains
\begin{equation}
\Tr\!\left[(L_{\nu,\mu\mu'}^{(+)})^\dagger L_{\nu,\mu\mu'}^{(+)}\right]
=
\gamma_\nu(\omega_{\mu\mu'}+\omega_c)\,
|\widetilde C_{\mu\mu'}|^2
\frac{n_{\max}(n_{\max}+1)}{2},
\end{equation}
while for the channel at frequency $\omega_{\mu\mu'}-\omega_c$,
\begin{equation}
\Tr\!\left[(L_{\nu,\mu\mu'}^{(-)})^\dagger L_{\nu,\mu\mu'}^{(-)}\right]
=
\gamma_\nu(\omega_{\mu\mu'}-\omega_c)\,
|\widetilde C_{\mu\mu'}|^2
\frac{n_{\max}(n_{\max}+1)}{2}.
\end{equation}
Thus, summing all first-order qubit-sector bath sideband channels gives
\begin{equation}
\begin{split}    
\sum_{k\in\nu}^{(1)}\Tr(L_k^\dagger L_k)
&=
\frac{n_{\max}(n_{\max}+1)}{2}
\sum_{\mu,\mu'=\pm}
|\widetilde C_{\mu\mu'}|^2\\
&\Big[
\gamma_\nu(\omega_{\mu\mu'}+\omega_c)
+\gamma_\nu(\omega_{\mu\mu'}-\omega_c)
\Big].
\end{split}
\label{eq:Tr_phonon_first_order_total}
\end{equation}

Accordingly, the first-order polaron correction to the qubit-sector-bath-induced
fidelity reduction is
\begin{equation}
\begin{split}
F_\nu^{(1)}
&=
\frac{t_g}{d+1}
\frac{n_{\max}(n_{\max}+1)}{2}
\sum_{\mu,\mu'=\pm}
|\widetilde C_{\mu\mu'}|^2\\
&\Big[
\gamma_\nu(\omega_{\mu\mu'}+\omega_c)
+
\gamma_\nu(\omega_{\mu\mu'}-\omega_c)
\Big].
\end{split}
\label{eq:F_first_order_phonon}
\end{equation}
\newpage
\bibliographystyle{apsrev4-2}
\bibliography{refs}

\end{document}